\documentclass[usenatbib]{mn2e}
\usepackage{graphicx}
\usepackage{amsmath}
\usepackage{times}
\usepackage{xspace}
\usepackage{color}
\usepackage{breqn}

\usepackage[normalem]{ulem}
\usepackage{enumitem}
\usepackage{float}
\usepackage{epsfig}
\usepackage{multirow}
\usepackage{epstopdf}
\usepackage{verbatim}
\usepackage{nicefrac}
\usepackage{url}
\usepackage{subfig}
\usepackage{bm}
\usepackage[table,usenames,dvipsnames]{xcolor}
\usepackage{amssymb}
\usepackage{booktabs}
\usepackage{threeparttable}
\usepackage{amsbsy}
\usepackage{amsfonts}

\usepackage{tabularx}
\newcolumntype{R}{>{\raggedleft\arraybackslash}X}

\newcommand{\sjaddress}{\url{https://github.com/sjoudaki/cfhtlens_revisited}\xspace}

\newcommand{\be}{\begin{equation}}
\newcommand{\ee}{\end{equation}}
\newcommand{\bea}{\begin{eqnarray}}
\newcommand{\eea}{\end{eqnarray}}

\newcolumntype{P}[1]{>{\centering\arraybackslash}p{#1}}

\newcommand{\nrun}{\ensuremath{dn_s/d\ln k}}

\newcommand{\halofit}{\textsc{halofit}\xspace}
\newcommand{\hmcode}{\textsc{hmcode}\xspace}
\newcommand{\athena}{\textsc{athena}\xspace}
\newcommand{\bpz}{\textsc{bpz}\xspace}
\newcommand{\cosmomc}{\textsc{CosmoMC}\xspace}
\newcommand{\plik}{\textsc{Plik}\xspace}
\newcommand{\camb}{\textsc{CAMB}\xspace}
\newcommand{\astac}{\textsc{ASTAC}\xspace}

\begin{document}

\voffset=-1.5cm
\hoffset=0.45cm

\title[CFHTLenS revisited]{CFHTLenS revisited: assessing concordance with Planck including astrophysical systematics}

\author[Joudaki et al.]{\parbox[t]{\textwidth}{Shahab Joudaki$^1$\thanks{E-mail: sjoudaki@swin.edu.au},
    Chris Blake$^1$, Catherine Heymans$^2$, Ami Choi$^2$, Joachim Harnois-Deraps$^3$, Hendrik Hildebrandt$^4$, Benjamin Joachimi$^5$, Andrew Johnson$^1$, Alexander Mead$^3$, David Parkinson$^6$, Massimo Viola$^7$, Ludovic van Waerbeke$^3$} \\ \\ $^1$ Centre for Astrophysics \&
  Supercomputing, Swinburne University of Technology, P.O.\ Box 218,
  Hawthorn, VIC 3122, Australia \\ $^2$ Scottish Universities Physics Alliance, 
  Institute for Astronomy, University of Edinburgh, Royal Observatory, Blackford
  Hill, Edinburgh, EH9 3HJ, U.K. \\ $^3$ Department of Physics and Astronomy, The University of British Columbia, 6224 Agricultural Road, Vancouver, B.C., V6T 1Z1, Canada \\ $^4$ Argelander Institute for Astronomy, University of Bonn, Auf dem Hugel 71, 53121 Bonn, Germany \\ $^5$ Department of Physics and Astronomy, University College London, London WC1E 6BT, U.K. \\ $^6$ School of Mathematics and Physics, University of Queensland, Brisbane, QLD 4072, Australia \\ $^7$ Leiden Observatory, Leiden University, Niels Bohrweg 2, 2333 CA Leiden, The Netherlands}

\pubyear{2016}
\date{\today}

\maketitle

\begin{abstract}
We investigate the impact of astrophysical systematics on cosmic shear cosmological parameter constraints from the Canada-France-Hawaii Telescope Lensing Survey (CFHTLenS), and the concordance with cosmic microwave background measurements by Planck. We present updated CFHTLenS cosmic shear tomography measurements extended to degree scales using a covariance calibrated by a new suite of N-body simulations. We analyze these measurements with a new model fitting pipeline, accounting for key systematic uncertainties arising from intrinsic galaxy alignments, baryonic effects in the nonlinear matter power spectrum, and photometric redshift uncertainties. We examine the impact of the systematic degrees of freedom on the cosmological parameter constraints, both independently and jointly. When the systematic uncertainties are considered independently, the intrinsic alignment amplitude is the only degree of freedom that is substantially preferred by the data. When the systematic uncertainties are considered jointly, there is no consistently strong preference in favor of the more complex models. We quantify the level of concordance between the CFHTLenS and Planck datasets by employing two distinct data concordance tests, grounded in Bayesian evidence and information theory. We find that the two data concordance tests largely agree with one another, and that the level of concordance between the CFHTLenS and Planck datasets is sensitive to the exact details of the systematic uncertainties included in our analysis, ranging from decisive discordance to substantial concordance as the treatment of the systematic uncertainties becomes more conservative. The least conservative scenario is the one most favored by the cosmic shear data, but it is also the one that shows the greatest degree of discordance with Planck. The data and analysis code are publicly available at \sjaddress.
\end{abstract}


\section{Introduction}
\label{Introduction}
\setcounter{footnote}{0}
\renewcommand{\thefootnote}{\arabic{footnote}}

The standard ${\Lambda}{\rm CDM}$ model of cosmology has been successful in describing the expansion history and growth of density perturbations throughout the Universe (e.g.~\citealt{anderson14, b14, planck15}). At the same time, it is facing challenges through our incomplete understanding of its main ingredients, namely the mechanism that is driving the current accelerated expansion and the dark matter (DM) that  constitutes most of the matter in the Universe (e.g.~\citealt{bertone05,copeland06,feng10,clifton12}). There are a range of late-time experimental techniques used to improve our understanding of the underlying cosmology of the Universe, such as supernova distances, baryon acoustic oscillations, galaxy cluster counting, and weak gravitational lensing, where lensing is considered to be one of the most promising as a result of its particular sensitivity to both structure formation and universal expansion (e.g.~\citealt{detf,JK12}). 

While weak lensing holds significant promise as a cosmological probe, the analyses of lensing datasets are still maturing. In particular, the optimism with weak lensing is predicated on overcoming the vast systematic uncertainties in both observations and theory. On the observational front, there are photometric redshift uncertainties (also denoted as `photo-z'; e.g.~\citealt{mhh06, htbj06,bh10,bonnett15}) and intrinsic alignments (IA) of galaxies (e.g.~\citealt{HS04, BK07,Joachimi11,joachimi15,ti15}), along with additive and multiplicative corrections to the lensing observables, for instance due to point spread function (PSF) anisotropies and 
shear miscalibration (e.g.~\citealt{hs03,htbj06,heymans12,mv12,bakm15}).

On the theoretical front, there are higher order correction terms in the lensing integral, for instance due to the Born approximation and lens-lens coupling, but these are negligible even for cosmic variance limited surveys (e.g.~\citealt{ch02,sc06,kh10,bbv10}). More crucially, lensing analyses need to account for reduced shear (e.g.~\citealt{dsw06,shapiro09,kh10}) and uncertainties in the modeling of the nonlinear matter power spectrum. 
The former, if neglected, may induce a bias in the cosmological parameter estimates that exceeds the parameter uncertainties in future surveys such as the Large Synoptic Survey Telescope~\citep{shapiro09}.
The latter is true even when assuming all of the matter is collisionless and a cosmological constant drives late-time universal acceleration, both analytically and with simulations 
(e.g.~\citealt{bcgs02,cs02,Smith03,Coyote4,mw15,Mead15}).

There are additional difficulties in modeling the matter power spectrum due to baryonic physics coming from star formation, radiative cooling, and feedback processes (e.g.~\citealt{white04,zk04,rudd08,Daalen11}). The modeling of the nonlinear matter power spectrum is also sensitive to extensions of the standard model, for example to include massive neutrinos (e.g.~\citealt{stt08,bird12,wagner12}), dark energy (e.g.~\citealt{mtc06,jch09,alimi10,Coyote4}), and modified gravity (e.g.~\citealt{stabenau06,zhao11,baldi14,hammami15}).

In this paper we present a methodical study of three key `astrophysical' systematic uncertainties affecting the lensing observables from the Canada-France-Hawaii Telescope Lensing Survey (CFHTLenS; \citealt{heymans12,hildebrandt12,erben13,miller13}) in the form of intrinsic galaxy alignments, baryonic effects in the nonlinear matter power spectrum, and photometric redshift uncertainties. 
In addition to these astrophysical uncertainties there are errors on the shear measurement itself which we calibrate through additive and multiplicative shear calibration corrections to the data as a function of galaxy size and signal-to-noise~\citep{heymans12, miller13}. In a 2D analysis, \citet{kilbinger13}~showed that the measured uncertainties in these corrections had a negligible impact on the cosmological constraints for CFHTLenS and so we do not consider `shear measurement' systematic uncertainties in our analysis (still true with tomography given comparable constraints on $\sigma_8 \Omega_{\mathrm m}^{0.5}$).

We account for the three key systematic uncertainties more comprehensively than previously, for instance by incorporating the halo-model based \hmcode \citep{Mead15,mead15code} to accurately include the baryonic signatures in the nonlinear matter power spectrum, and by allowing for a possible luminosity and redshift dependence of the intrinsic alignments (in addition to the amplitude dependence).                            
We also account for biases in the measured redshift distribution for each tomographic bin, both by considering random shifts around the fiducial distributions, and by considering systematic shifts to the distributions following the analysis of source-lens cross-correlations in \citet{choi15}.
We consider these systematic uncertainties both independently and jointly, and ask if the data 
favors any of the additional degrees of freedom. 
For the purposes of model selection, we use the deviance information criterion (DIC;~\citealt{spiegelhalter02}), and complement with calculations of the Bayesian evidence (e.g.~\citealt{mnest1, trotta08}). These statistical tools are discussed in Section~\ref{modsec}.

We also strive to improve our understanding of the `discordance' in the cosmological constraints from the cosmic shear and cosmic microwave background (CMB) datasets of CFHTLenS and Planck (e.g.~\citealt{planck13, maccrann15, planck15, raveri15, grandis15}). 
To achieve this, we take a methodical approach. We begin with the minimal scenario where no systematic uncertainties are included in the analysis of CFHTLenS, and examine the potential dependence of the results to the choice of cosmological priors. We then consider a whole series of scenarios where the key systematic uncertainties are included independently and jointly, both with informative priors and with non-informative priors. We employ data concordance tests based on the Bayesian evidence and deviance information criterion, and find that the level of discordance between the CFHTLenS and Planck datasets is sensitive to the assumptions made on the level of systematic uncertainties in the CFHTLenS measurements, such that increasingly conservative scenarios show an increasing degree of concordance between the datasets.

In addition to the comprehensive account of the systematic uncertainties, we update the CFHTLenS measurements first presented in the 6-bin tomographic analysis of Heymans et al.~(2013; also denoted H13). As described in Section~\ref{measlab}, we divide the source galaxies into 7 tomographic bins with redshift ranges that allow us to more optimally account for the overlap with spectroscopic surveys in forthcoming analyses. We moreover extend the angular coverage of the measurements from $[1, 50]$ arcmins in \citet{Heymans13} to $[1, 120]$ arcmins in this work, owing to the increased box size of the new N-body simulations used to determine the data covariance matrix (\citealt{hdw15}; described in Section~\ref{covlab}). Thus, instead of the original 5 angular bins, we now have 7 angular bins in the aforementioned range. 

In Section~\ref{Methodology}, we present the theoretical basis of our work, along with our updated CFHTLenS measurements and covariance matrix estimation from N-body simulations. In Section~\ref{results}, we explore the impact of the systematic uncertainties on the cosmological constraints, independently and jointly. We examine whether the new degrees of freedom are favoured by the data, and investigate the level of concordance between the CFHTLenS and Planck datasets.
In Section~\ref{conclusions}, we conclude with a discussion of our results.

\section{Methodology}
\label{Methodology}

We give an overview of the theory associated with weak gravitational lensing and intrinsic galaxy alignments. We discuss our new fitting pipeline, and the methods by which we account for photometric redshift uncertainties and baryonic uncertainties in the nonlinear matter power spectrum. We then proceed to describe our updated CFHTLenS measurements and covariance matrix considering 7 tomographic bins. We do not include additional degrees of freedom for the additive and multiplicative shear calibration corrections, but incorporate these directly in our data.

\subsection{Theory}

\subsubsection{Weak lensing observables}
We follow the standard approach in computing the weak lensing observables (e.g.~\citealt{BS01}), 
in the form of the 2-point shear correlation functions,
\begin{equation}
\xi_{\pm}^{ij}(\theta)_{\rm GG} = \frac{1}{2\pi}\int d\ell \,\ell \,C_{\rm GG}^{ij}(\ell) \, J_{\pm}(\ell \theta) \, , 
\label{eqn:xipmgg}
\end{equation}
defined at angle $\theta$, where $C_{\rm GG}^{ij}(\ell)$ is the convergence power spectrum for tomographic bin combination $\{i,j\}$ at angular wavenumber $\ell$, and $J_{\pm}$ are the zeroth (+) and fourth (-) order Bessel functions of the first kind. 
Given seven tomographic bins, $i$ and $j$ both run from 1 to 7, such that there are 28 independent combinations.
Using the Limber approximation~(\citealt{Limber}, also see~\citealt{LA08}), the convergence power spectrum is then obtained as a weighted integral over the matter power spectrum,
\begin{equation}
C_{\rm GG}^{ij}(\ell) = \int_0^{\chi_{\rm H}} d\chi \, 
\frac{q_i(\chi)q_j(\chi)}{[f_K(\chi)]^2} \, P_{\delta\delta} \left(\frac{\ell+1/2}{f_K(\chi)},\chi \right),
\label{eqn:ckk} 
\end{equation}
where $\chi$ is the comoving distance, $\chi_{\rm H}$ is the comoving horizon distance, $f_K(\chi)$ is the comoving angular diameter distance, $P_{\delta\delta}$ is the matter power spectrum, 
and the geometric weight $q_i(\chi)$ in tomographic bin $i$ is given by
\begin{equation}
q_i(\chi) = \frac{3 H_0^2 \Omega_{\mathrm m}}{2c^2} \frac{f_K(\chi)}{a(\chi)}\int_\chi^{\chi_{\rm H}}\, d\chi'\ n_i(\chi') \frac{f_K(\chi'-\chi)}{f_K(\chi')}.
\label{eqn:qi} 
\end{equation}
Here, $a(\chi)$ is the scale factor, $c$ is the speed of light, $H_0$ is the Hubble constant, $\Omega_{\mathrm m}$ is the present matter density, and $n_i(\chi)$ encodes the source galaxy distribution in a given tomographic bin, normalized to integrate to unity.

\subsubsection{Intrinsic galaxy alignments}
\label{iatheory}

We further extend our theory to account for intrinsic galaxy alignments~\citep{HS04,BK07,Joachimi11}, originating from correlations of intrinsic ellipticities of galaxies with each other and with the shear of background sources. These intrinsic-intrinsic (II) and shear-intrinsic (GI) terms enter the observed correlation function, such that
\begin{equation}
\xi_{\pm}^{ij}(\theta)_{\rm obs} = \xi_{\pm}^{ij}(\theta)_{\rm GG} + \xi_{\pm}^{ij}(\theta)_{\rm II} + \xi_{\pm}^{ij}(\theta)_{\rm GI}.
\label{eqn:xipmtotal} 
\end{equation}
Here, the II and GI terms are defined as in equation~(\ref{eqn:xipmgg}), except $C_{\rm GG}^{ij}$ is correspondingly replaced by $C_{\rm II}^{ij}$ and $C_{\rm GI}^{ij}$. 
{Following \citet{BK07} in using the nonlinear matter power spectrum within the linear theory of \citet{HS04}, we express the II term:}
\begin{equation}
C_{\rm II}^{ij}(\ell) = \int_0^{\chi_{\rm H}} d\chi \, 
\frac{n_i(\chi)n_j(\chi)F_i(\chi)F_j(\chi)}{[f_K(\chi)]^2} \, P_{\delta\delta} \left( \frac{\ell+1/2}{f_K(\chi)},\chi \right),
\label{eqn:xipmii}
\end{equation}
and the GI term:
\begin{dmath}
C_{\rm GI}^{ij}(\ell) = \int_0^{\chi_{\rm H}} d\chi \frac{q_i(\chi)n_j(\chi)F_j(\chi)}{[f_K(\chi)]^2}P_{\delta\delta} \left( \frac{\ell+1/2}{f_K(\chi)},\chi \right) + 
\int_0^{\chi_{\rm H}} d\chi \frac{n_i(\chi)F_i(\chi)q_j(\chi) }{[f_K(\chi)]^2} P_{\delta\delta} \left({\frac{\ell+1/2}{f_K(\chi)},\chi}\right).
\label{eqn:xipmgi}
\end{dmath}
We allow for an unknown amplitude $A$ along with a possible redshift ($z$) and luminosity ($L$) dependence via $\eta$ and $\beta$, respectively, in defining
\begin{equation}
F_i(\chi) = - A C_1 \rho_{\rm cr} \frac{\Omega_{\mathrm m}}{D(\chi)} \left({1+z(\chi)} \over {1+z_0}\right)^\eta \left({L_i \over L_0}\right)^\beta ,
\label{eqn:fz}
\end{equation}
in accordance with~\citet{Joachimi11}, where $\rho_{\rm cr}$ is the critical density at present, $D(\chi)$ is the linear growth factor normalized to unity at present, the normalization constant $C_1 = 5 \times 10^{-14} \, h^{-2} M_\odot^{-1} {\rm Mpc}^3$, $z_0 = 0.3$ is an arbitrary pivot redshift, and $L_0$ is the pivot luminosity corresponding to an absolute r-band magnitude of -22. We determine the luminosities by averaging the individual galaxy luminosities (calculated as $10^{-0.4M}$ for each galaxy, where $M$ is the absolute magnitude), weighted by the $\emph{lens}$fit weights 
(defined in Section~\ref{measlab}), 
giving us an effective $L_i/L_0 = (0.017, 0.069, 0.15, 0.22, 0.36, 0.49, 0.77)$ for our seven tomographic bins.

\begin{figure*}
\begin{center}
\resizebox{14cm}{!}{\rotatebox{270}{\includegraphics{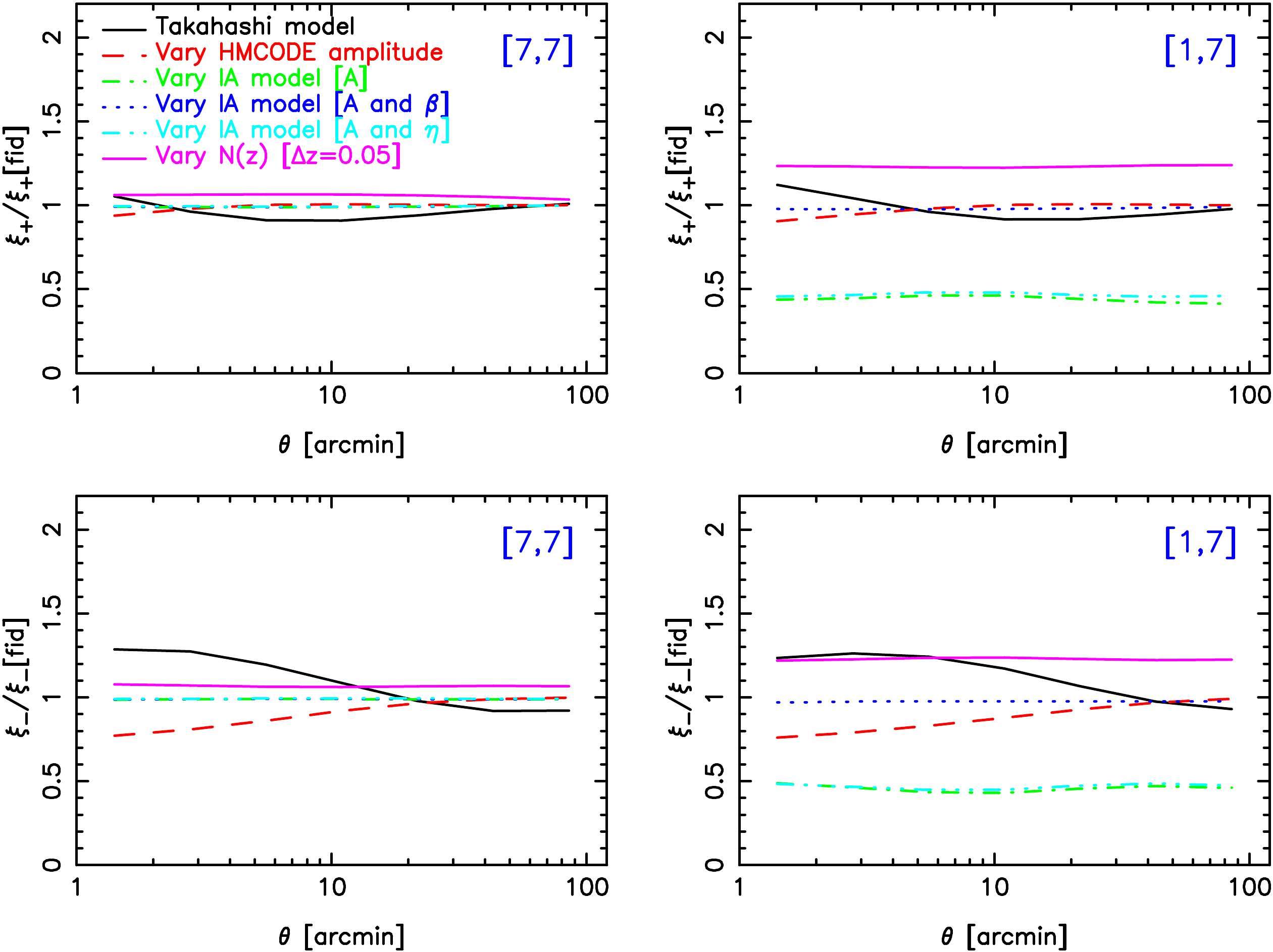}}}
\end{center}
\caption{The ratio of shear correlation functions for tomographic bin combinations \{1,7\} and \{7,7\}, taken with respect to \hmcode with feedback amplitude $\log B = 0.496$, defined in equation~(\ref{eq:cm}), including no systematic uncertainties (denoted as $\xi_\pm[{\rm fid}]$). For consistency, we fix the underlying cosmology to that of the best-fit cosmology of this `fiducial' case. We allow for the \citet{Takahashi12} version of \halofit (solid black), \hmcode with $\log B = 0.3$ (dashed red), intrinsic alignments with $\{A, \eta, \beta\} = \{1,0,0\}$ (dot-dashed green), intrinsic alignments with $\{A, \eta, \beta\} = \{1,0,1\}$ (dotted blue), intrinsic alignments with $\{A, \eta, \beta\} = \{1,1,0\}$ (dot-dashed cyan), and photometric redshift uncertainties where all bins are positively perturbed by ${\Delta}z = 0.05$ (solid pink). 
The $\log B$ value of 0.496 corresponds to the DM-only case, while $\log B = 0.3$ agrees with the AGN case of the OWL simulations.
The parameters $\{A, \eta, \beta\}$ refer to the intrinsic alignment amplitude, redshift dependence, and luminosity dependence, respectively, of the IA model defined in equation~(\ref{eqn:fz}), while the photo-z shifts are defined in equation~(\ref{eqn:photoz}). The IA model with $\{A, \eta, \beta\} = \{1,0,1\}$ lies along the unity line because the luminosity $L/L_0 < 1$ in each tomographic bin, such that a positive value of $\beta$ suppresses the IA signal (analogously, a negative value of $\eta$ for the redshift dependence would have a similar effect).
}
\label{figmod}
\end{figure*}

\subsubsection{Photometric redshift uncertainties}
\label{photheory}

We account for uncertainties in the photometric redshift estimation by allowing the redshift distribution in each tomographic bin to shift along the redshift axis by an amount ${\Delta}z_i$, such that
\begin{equation}
n_i^{\rm theory}(z) = n_i^{\rm obs}(z - {\Delta}z_i),
\label{eqn:photoz}
\end{equation}
where $n_i^{\rm obs}$ is the observed redshift distribution.
This is consistent with the approach used in \citet{dessv}.
As a minor caveat, as we do not integrate below the minimum redshift of $z_{\rm min}$ = 0.03 for the fiducial redshift distributions (the necessity of $z_{\rm min} > 0$ is because $z=0$ would correspond to wavenumber $k = \infty$), we take a consistent approach and continue to neglect the same lowest end of the redshift distributions when shifted to higher redshifts. A physical interpretation of this would be to consider these outliers as stars at $z = 0$. Since we consider 7 tomographic bins, this introduces an additional 7 nuisance parameters that we marginalize over in our analysis, with either uniform or Gaussian priors (e.g.~via cross-correlations with an overlapping spectroscopic sample), as further discussed in Section~\ref{photsec}.

\subsubsection{Baryonic uncertainties in the nonlinear matter power spectrum:~\hmcode}
\label{barytheory}

We account for baryonic uncertainties in the nonlinear matter power spectrum by incorporating \hmcode in \citet{Mead15}\footnote{\texttt{https://github.com/alexander-mead/hmcode}} as a separate parallelized module in \cosmomc (\citealt{Lewis:2002ah}\footnote{\texttt{http://cosmologist.info/cosmomc/}}). By introducing physically motivated free parameters in the halo-model formalism, and calibrating these to the Coyote N-body dark matter simulations (\citealt{Coyote4} and references therein), \hmcode is able to describe the power spectrum to marginally improved accuracy in comparison to the latest incarnation of \halofit~\citep{Smith03, Takahashi12}. 

However, by further calibrating to the OverWhelmingly Large (OWL) Simulations \citep{Schaye10,Daalen11}, the main benefit of \hmcode is its capacity to account for baryonic effects in the matter power spectrum on nonlinear scales, for example due to star formation, radiative cooling, and AGN feedback. This is achieved by modifying 
parameters that govern the internal structure of halos. 
For example, AGN feedback blows gas out of halos, which makes them less concentrated.~\citet{Mead15} found that acceptable fits could be made to the OWL simulations using a halo model with less concentrated halos in the one-halo term. Conversely the two-halo term is unchanged because feedback only affects small scales.
Thus, \hmcode modifies the relationship between halo concentration $c$ and halo mass $M$, such that
\begin{equation}
c(M,z)=B\frac{1+z_{\rm f}}{1+z} , 
\label{eq:cm}
\end{equation} 
where $z_{\rm f}$ is the halo formation redshift as a function of halo mass and $B$ is a free parameter that we can marginalize over in our analysis. 
\hmcode can also change the halo density profile via the halo bloating parameter $\eta_{\hmcode}$ to account for baryonic effects. However, in \citet{Mead15} it was shown that substantial degeneracy exists between $\eta_{\hmcode}$ and $B$ and that these two parameters can be linearly related to provide a one-parameter baryonic feedback model. We use this prescription and $\eta_{\hmcode}$ therefore does not contribute an additional degree of freedom in our Markov Chain Monte Carlo (MCMC) analysis.

For scales $k < 10~h^{-1} {\rm Mpc}$, \hmcode produces a nonlinear matter power spectrum that accounts for baryonic physics (REF, DBLIM, and AGN cases of the OWL simulations, described in \citealt{Daalen11}) accurately at the level of a few percent. \hmcode's ability to accurately model the nonlinear matter power spectrum including baryons with a single parameter can be contrasted with the fitting formula in \citet{hwvh15}, which employs 15 free parameters to achieve the same outcome with similar precision 
(also see \citealt{kohlinger15}, which use the same prescription as in \citealt{hwvh15} but only marginalize over a single parameter, and \citealt{maccrann15} for a similar single-parameter marginalization approach based on the AGN case of the OWL simulations).
 
While \hmcode is calibrated to $k < 10~h^{-1} {\rm Mpc}$, it agrees with the matter power spectrum from Takahashi et al (2012) at the 10\% level for $k < 100~h^{-1} {\rm Mpc}$ (DM-only case). Even if we assume that \hmcode miscalibrates the matter power spectrum (including baryons) for $k > 10~h^{-1} {\rm Mpc}$ by a factor of 2 for every decade in wavenumber beyond $k = 10~h^{-1} {\rm Mpc}$, it would bias the lensing correlations functions by at most 1\% for $\xi_+$ and at the sub-percent level for $\xi_-$ for the angular scales considered in this work (less for $\xi_-$ than for $\xi_+$ due to our angular cuts; checked for the case without tomography). Given the statistical power of our data 
(described in Section~\ref{measlab}), 
the accuracy of \hmcode is therefore sufficient for our purposes, and it forms an important component of our new pipeline.

\subsubsection{New \cosmomc module for WL analyses with systematic uncertainties}
\label{modtheory}

In order to account for the systematic uncertainties coming from intrinsic alignments of galaxies, photometric redshift uncertainties, and baryonic effects in the nonlinear matter power spectrum, we have developed a new module in \cosmomc (in the language of Fortran 90). The module is independent from previous lensing modules, and accounts for the systematic uncertainties following the prescriptions in  Sections~\ref{iatheory} to \ref{barytheory}.

The new module allows the user to choose the integration method for the \{GG, II, GI\} spectra with one of two distinct methods, either with trapezoidal integration or with Romberg integration. We have internally parallelized the code, 
which with a single eight-core Intel Xeon E5-2660 processor at 2.2 GHz can calculate the likelihood for a single cosmology, considering 6 tomographic bins and including intrinsic alignments (i.e.~GG, II, and GI), in 0.078 seconds when using \halofit for the nonlinear matter power spectrum.
Since our module is parallelized the speed would continue to show some improvement with further cores.
This can be compared to the existing default lensing module in \cosmomc which with the same resources calculates the likelihood for a single cosmology, considering 6 tomographic bins and without intrinsic alignments (i.e.~only GG), in 0.33 seconds.\footnote{In the development of this module, we 
verified our results by comparing against a completely independent implementation available to the collaboration (Joachimi) and to the default CosmoMC lensing module. The lensing observables calculated with our new module agree well with those calculated with the collaboration's independent code. There were however discrepancies in the convergence power spectrum with the default CosmoMC lensing module (which seem to be caused by insufficiently accurate integration in the default CosmoMC lensing module; here we only checked GG as the default CosmoMC lensing module does not account for intrinsic alignments). We find the discrepancies to be negligible at the level of the parameter constraints due to the sufficiently weak statistical power of current data.}
We note that these are the speeds of only the respective modules, i.e. the numbers do not account for the time it takes the Boltzmann code \camb~(\citealt{LCL}\footnote{\texttt{http://camb.info}}) to compute the matter power spectrum which is fed into both modules.

As we have incorporated \hmcode as a separate parallelized module in \cosmomc, at each new cosmology, the lensing module internally provides \hmcode the linear matter power spectrum obtained from a modified version of CAMB in a ($k,z$)-array and obtains from it the nonlinear matter power spectrum at the same ($k,z$) values in return. Using the same processor, this transition between linear to nonlinear power spectrum takes 0.4 seconds for a ($k,z$)-array that is sufficiently dense for our lensing calculation. While the computation of the nonlinear matter power spectrum with \hmcode is slower than the computation with \halofit, it allows us to account for the baryonic effects on nonlinear scales. However, when nonlinear baryonic effects are not considered, the agreement between \hmcode and \halofit is sufficiently close that either one could be used.

In Fig.~\ref{figmod}, we show the impact of the different systematic degrees of freedom on $\{\xi_+, \xi_-\}$ for the tomographic bin combinations \{1,7\} and \{7,7\}. As expected, we find that the difference between the \halofit and \hmcode prescriptions enters the observables on smaller angular scales. Moreover, varying the \hmcode feedback amplitude leaves an imprint on the observables that increases with smaller angular scales. Analogously, the imprint of the different nonlinear prescriptions (both with and without baryons) is larger for $\xi_-$ than for $\xi_+$, due to the greater sensitivity of $\xi_-$ to nonlinear scales in the matter power spectrum for a given angular scale. 
Meanwhile, the impact of intrinsic alignments and shifts in the photometric redshift distributions seem to be strongest in the cross-bins and fairly independent of angular scale. 
This illustrates the usefulness of these bins in constraining the intrinsic alignment model and deviations from the fiducial photometric redshift distributions.

We have further extended our module to account for joint analyses of overlapping observations of cosmic shear, galaxy-galaxy lensing, and large-scale structure measured through clustering multipoles,
including the full covariance, as part of our efforts to constrain modified gravity and neutrino physics. 
Along with the new CFHTLenS measurements, we are releasing our code (both cosmic shear and \hmcode modules in CosmoMC) pertaining to the calculations presented in this paper at \sjaddress. We will be releasing our full code as part of an upcoming paper (Joudaki et al. in prep).

Lastly, we note that our module is currently independent from the CosmoSIS platform \citep{cosmosis}, which combines a range of disparate codes into a single framework for cosmological parameter estimation. There are no technical obstacles to prevent our module from being incorporated into CosmoSIS in the future.

\subsubsection{Baseline configurations}
\label{baseconfig}

In our analysis, we always include the `vanilla' parameters, given by $\left\{{\Omega_{c}h^2, \Omega_{b}h^2, \theta_{\rm MC}, n_{s}, \ln{(10^{10} A_{s})}}\right\}$, which represent the cold dark matter density, baryon density, approximation to the angular size of the sound horizon (in \cosmomc), scalar spectral index, and amplitude of the scalar spectrum, respectively.
We note that `$\ln$' refers to the natural logarithm, while we take `$\log$' to refer to logarithms with base 10. From these parameters, one can derive the Hubble constant $H_0$ (also expressed as $h$ in its dimensionless form) and standard deviation of the present linear matter density field on scales of $8~h^{-1} {\rm Mpc}$ (denoted by $\sigma_8$).
We impose uniform priors on these cosmological parameters, as discussed in Section~\ref{results}.

In this baseline $\Lambda$CDM model, we include 3 massless neutrinos, such that the effective number of neutrinos $N_{\rm eff} = 3.046$ (we have checked that our results are not significantly affected by the approximation of zero mass, as compared to the minimal mass of the normal hierarchy of 0.06 eV). For the primordial fraction of baryonic mass in helium, $Y_p$, we determine the quantity as a function of $\{N_{\mathrm{eff}}, \Omega_b h^2\}$ in a manner consistent with Big Bang Nucleosynthesis (BBN; see equation~1 in \citealt{sj13}).
Moreover, we consistently enforce the strong inflation prior on the curvature and running of the spectral index, such that $\{\Omega_k \equiv 0, \nrun \equiv 0\}$. Thus, with flatness enforced, $\Omega_{\mathrm m} > 1$ implies $\Omega_\Lambda < 0$. Lastly, with no running of the spectral index, we define the primordial scalar power spectrum,
\begin{equation}
\ln P_s (k) = \ln A_s + (n_s - 1) \ln(k/k_{\rm pivot}),
\label{eq:plaw}
\end{equation}
where both $A_s$ and $n_s$ are defined at the pivot wavenumber $k_{\rm pivot}$.

In order to determine the convergence of our MCMC chains, we use the \citet{Gelman92} $R$ statistic, where $R$ is defined as the variance of chain means divided by the mean of chain variances. 
Our runs are stopped when the conservative limit $(R - 1) < 2 \times 10^{-2}$ is reached, 
and we have checked that further exploration of the tails does not change our results.

\subsubsection{Model selection and dataset concordance}
\label{modsec}

We define the best-fit effective $\chi^2$, via ${\chi^2_{\rm eff}(\hat{\theta})}  = -2 \ln {\mathcal{L}}_{{\rm max}}$, where ${\mathcal{L}}_{\rm max}$ is the maximum likelihood of the data given the model, $\theta$ is the vector of varied parameters, and hat denotes the maximum likelihood point. When quoting $\chi^2_{\rm eff}$ without specifying $\theta$, we implicitly assume $\theta = \hat{\theta}$. The reduced $\chi^2$ is then given by $\chi^2_{\rm red} =\chi^2_{\rm eff} / \nu$, where $\nu$ is the number of degrees of freedom.
Given two separate models, where $\Delta\chi^2_{\rm eff} > 0$, we interpret the model with the higher value of $\chi^2_{\rm eff}$ to be associated with a lower probability of drawing the data at the maximum likelihood point, by a factor given by $\exp({-\Delta\chi^2_{\rm eff}/2})$. 
For reference, a difference of 10 in $\chi^2_{\rm eff}$ between two models would correspond to a probability ratio of 1 in 148, and therefore constitute strong preference for the more probable model. 

When considering the relative performance of two distinct models,
it is valuable to compute the Deviance Information Criterion (DIC; \citealt{spiegelhalter02}, also see \citealt{ktp06}, \citealt{liddle07}, \citealt{trotta08}, and \citealt{spiegelhalter14}), obtained from the Kullback-Leibler divergence or relative information entropy (\citealt{kl51}). 
We do not use the Akaike Information Criterion~\citep{akaike}, 
which follows from an approximate minimization of the Kullback-Leibler divergence and does not account for unconstrained directions in parameter space (e.g.~\citealt{liddle07,trotta08}). 
We also do not consider the Bayesian Information Criterion~\citep{bic}, as it is not grounded in information theory and instead follows from a Gaussian approximation to the Bayesian evidence (e.g.~\citealt{liddle07,trotta08}).
In practice, we compute
\begin{equation}
{\rm{DIC}} \equiv {\chi^2_{\rm eff}(\hat{\theta})} + 2p_D, 
\label{dicdef}
\end{equation}
where $p_D = \overline{\chi^2_{\rm eff}(\theta)} - {\chi^2_{\rm eff}(\hat{\theta})}$ is the `Bayesian complexity' and the bar denotes the mean taken over the posterior distribution~(\citealt{spiegelhalter02}). 
The Bayesian complexity is a measure of the effective number of parameters, 
and acts as a penalty against more complex models.
Instead of the maximum likelihood point, the Bayesian complexity and DIC are also commonly evaluated at the posterior mean or median of the cosmological parameters (e.g.~\citealt{spiegelhalter02, trotta08}).
The main limitation of the DIC is the $\chi^2_{\rm eff}$ of the posterior mean can sometimes not be a good fit to the data in multimodal distributions, and the alternative (the best fit $\chi^2_{\rm eff}$) is a somewhat arbitrary choice. Moreover, the DIC uses the data effectively twice (in that the `penalty factor' also depends on the data), and its use of a point estimate can be stochastically affected by the data. Finally, 
beyond brute force no efficient and accurate method has been developed for computing the errors of the DIC estimates.

For two models with the same complexity, the difference in their DIC values is the same as the difference in their respective $\chi^2_{\rm eff}$ values. 
Analogous to the $\chi^2_{\rm eff}$ scenario, a difference of 10 in DIC between two models constitutes strong preference in favor of the model with the lower DIC estimate,
while a difference of 5 in DIC between two models constitutes moderate preference in favor of the model with the lower DIC estimate.
When the difference in DIC between two models is even smaller, the statistic only weakly favors one model over the other.
In comparing an extended model with a reference model, we take negative values of $\Delta{\rm DIC}$ to indicate a preference in favor of the extended model as compared to the reference model.

We complement our DIC analysis by using the nested sampling algorithm CosmoChord \citep{pchord1,pchord2} to compute the Bayesian evidence (with additional runs using MultiNest to ensure consistency in the results; \citealt{mnest1,mnest2,mnest3}). The evidence is given by 
the average of the likelihood under the prior for a given model,
\begin{equation}
{\mathcal{Z}} = \int d^n{\bm{\theta}}~{\mathcal{L}}({\bm{\theta}}) \pi({\bm{\theta}}),
\label{eqevidence}
\end{equation}
where $n$ encapsulates the dimensionality of the parameter space 
and $\pi({\bm{\theta}})$ is the prior given the vector of parameters ${\bm{\theta}}$ (e.g.~\citealt{mnest1,trotta08}). 
Equation~(\ref{eqevidence}) tells us that the evidence is larger for a simpler theory with a compact parameter space, unless it is significantly worse at explaining the data as compared to a more complicated theory. 
For model selection purposes, we also compute the Bayes factor (e.g.~\citealt{mnest1,trotta08}), given by the evidence ratio for two specific models, denoted by ${\mathcal{Z}_0}$ and ${\mathcal{Z}_1}$:
\begin{equation}
\mathcal{B}_{01} \equiv {{\mathcal{Z}_0}/{\mathcal{Z}_1}}.
\label{bayesfac}
\end{equation}
For a scenario in which the prior probabilities of the two models are equal, the Bayes factor encapsulates the posterior odds, such that the data favors model 0 as compared to model 1 when the Bayes factor is greater than unity, and vice versa. Alternatively, the Bayes factor can be thought of as the change to the prior odds given the data.

From the evidence calculations, we can further construct a measure of the concordance between two datasets $D_1$ and $D_2$, given by
\begin{equation}
{\mathcal{C}}(D_1, D_2) \equiv {{{\mathcal{Z}}(D_1 \cup D_2)} \over {{\mathcal{Z}}(D_1) {{\mathcal{Z}}(D_2)}}}, 
\label{eveqn}
\end{equation}
where ${{\mathcal{Z}}(D_1 \cup D_2)}$ is the joint evidence of the two datasets \citep{mrs06,raveri15}. 
Thus, $\log \mathcal{C}$ is positive when there is concordance between the two datasets, such that the joint evidence is larger than the product of the individual evidences, and similarly $\log \mathcal{C}$ is negative when there is discordance between the two datasets. We will use this concordance test to  better assess the potential degree of tension between the updated CFHTLenS and Planck measurements. In this pursuit, we also introduce an analogous but more easily calculable quantity from the DIC estimates:
\begin{equation}
{\mathcal{I}}(D_1, D_2) \equiv \exp\{{-{\mathcal{G}}(D_1, D_2)/2}\}, 
\label{diceqn}
\end{equation}
where
\begin{equation}
{\mathcal{G}}(D_1, D_2) = {{{\rm{DIC}}(D_1 \cup D_2)} - {{\rm{DIC}}(D_1) - {{\rm{DIC}}(D_2)}}},
\end{equation}
and ${{\rm{DIC}}(D_1 \cup D_2)}$ is the joint DIC of the two datasets. 
We expect this quantity to diagnose separation or congruence between posterior distributions, through measurement of the relative entropy of one distribution with respect to the other. To describe it in terms of a Gaussian example, if two datasets that agree are added together, we would expect the joint likelihood to have a larger ${\chi^2_{\rm eff}(\hat{\theta})}$ (since there are more data points), roughly equivalent to the sum of the individual ${\chi^2_{\rm eff}}$, but the same Bayesian complexity for both, leading to an overall negative value for $\mathcal{G}(D1, D2)$ (since the complexity factor is applied twice), and so to a large $\mathcal{I}$. However, if the two datasets do not agree, there will be a much larger ${\chi^2_{\rm eff}}$ than the sum of the individual ${\chi^2_{\rm eff}}$, and this will not be balanced enough by the change in complexity, since the different parameters will not be measured well. In this case, the overall value for $\mathcal{G}(D1, D2)$ will be positive, leading to a small $\mathcal{I}$.

Thus, analogous to the evidence scenario, $\log \mathcal{I}$ is constructed such that there is concordance between the datasets when it is positive, and discordance between the datasets when it is negative. For an independently developed concordance test based on the Kullback-Leibler divergence, see \citet{seehars15}. We further assess the degree of concordance or discordance by employing Jeffreys' scale (\citealt{jeffreys}, also see \citealt{kr95}), such that values for $\log \mathcal{C}$ and $\log \mathcal{I}$ in excess of $\pm 1/2$ are `substantial', values in excess of $\pm 1$ are `strong', and values in excess of $\pm 2$ are `decisive' (where this last case corresponds to a probability ratio in excess of 100).

\subsection{Measurements}
\label{measlab}

In this section, we introduce the CFHTLenS dataset and new measurements
used in our cosmology analysis. The CFHTLenS\footnote{\texttt{http://www.cfhtlens.org}} 
is a deep multi-colour survey optimized for weak lensing analyses, 
based on data from the Canada-France-Hawaii Telescope (CFHT) Legacy Survey in five optical bands $u^* g' r' i' z'$, using the 1 deg$^2$ camera MegaCam. The Wide Survey data analyzed in this study span four fields W1, W2, W3 and W4, which together cover 154~deg$^2$. 

Galaxy ellipticity components $(e_1, e_2)$ for each source, together with an approximately optimal
inverse-variance weight $w^s$, are determined by the Bayesian
model-fitting software {\it lens}fit \citep{miller13}.
Photometric redshifts are derived from PSF-matched photometry
\citep{hildebrandt12} using the Bayesian photometric redshift
code \bpz \citep{benitez2000}, which also returns a full 
redshift probability distribution $p_{\rm BPZ}(z)$, with peak $z_B$ for each source.
The survey pointings have been subjected to a stringent
cosmology-independent systematic-error analysis \citep{heymans12}, as 
a result of which a subset of around $25\%$ of the
pointings have been flagged as possessing potentially significant systematic errors,
and are excluded from our analysis. We applied additive shear
calibration corrections to the measured ellipticities, and
multiplicative shear calibration corrections to the cosmic shear
measurements (following \citealt{heymans12} and \citealt{miller13}).
We only retain unmasked sources for our analysis.

\begin{figure}
\begin{center}
\resizebox{8cm}{!}{\rotatebox{270}{\includegraphics{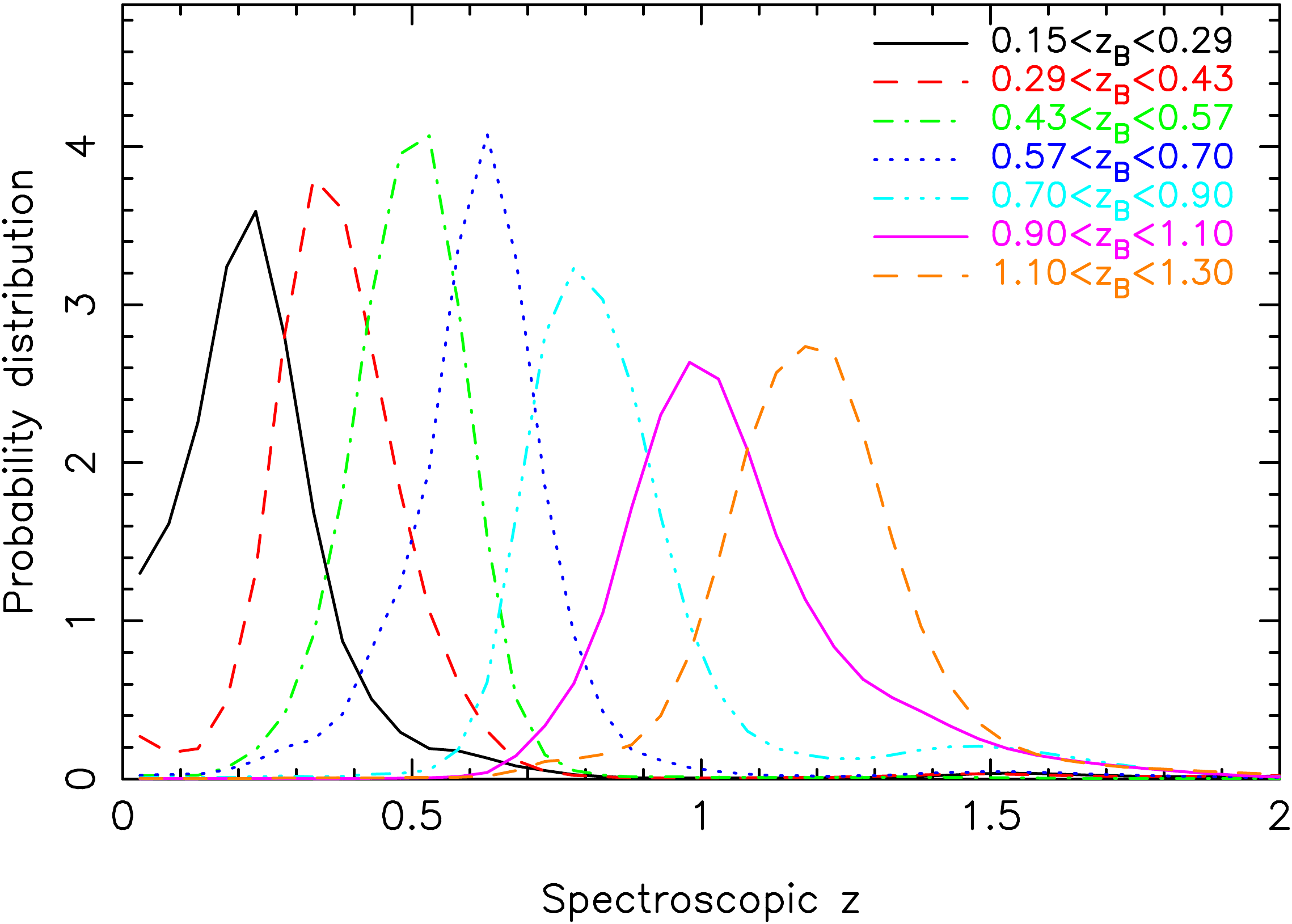}}}
\end{center}
\caption{Stacked \bpz redshift probability distributions for
  CFHTLenS, weighted by the {\it lens}fit weights, in the seven
  tomographic photo-$z$ bins used in our analysis.}
\label{fignz}
\end{figure}

We perform cosmic shear tomography by dividing the sources according
to $z_B$ into $N_t = 7$ tomographic bins with ranges $0.15-0.29,
0.29-0.43, 0.43-0.57, 0.57-0.70, 0.70-0.90, 0.90-1.10, 1.10-1.30$.
This choice represents a slight alteration from the original CFHTLenS
tomographic analysis \citep{Heymans13}, which divided the range
$0.2 < z_B < 1.3$ into 6 tomographic bins.  This modification was
motivated by the overlapping spectroscopic datasets now available due
to the Baryon Oscillation Spectroscopic Survey (BOSS, \citealt{anderson14}), which are conveniently split into redshift ranges $0.15-0.43$ (LOWZ sample) and $0.43-0.70$ (CMASS sample). 
This spectroscopic redshift data may be used to include galaxy-galaxy
lensing and redshift-space distortion statistics in the analysis (with
appropriate covariance), and further calibrate the photometric
redshifts through cross-correlation (e.g.~\citealt{choi15}). Figure \ref{fignz} displays the
stacked \bpz redshift probability distributions, weighted by the
{\it lens}fit weights, for each tomographic source bin. A spline
function of these measurements is used as the model source redshift
distribution in our cosmology-fitting pipeline.

The effective source density for lensing analyses is defined by
\begin{equation}
n_{\rm eff} = \frac{1}{A_{\rm eff}} \, \frac{ \left( \sum_i w^s_i
  \right)^2 }{\sum_i (w^s_i)^2} ,
\label{eqsourcedens}
\end{equation}
where $A_{\rm eff}$ is the effective (unmasked) area. In the range
$0.15 < z_B < 1.3$ used in our study, the values derived for the four
survey regions \{W1, W2, W3, W4\} are $n_{\rm eff} = \{10.9, 9.9,
11.0, 10.4\}$ arcmin$^{-2}$ for unmasked areas $\{42.9, 12.1, 26.1,
13.3\}$ deg$^2$; for the whole sample we find $n_{\rm eff} = 10.7$
arcmin$^{-2}$ over $A_{\rm eff} = 94$~deg$^2$.

\begin{figure*}
\begin{center}
\resizebox{17.5cm}{!}{\rotatebox{270}{\includegraphics{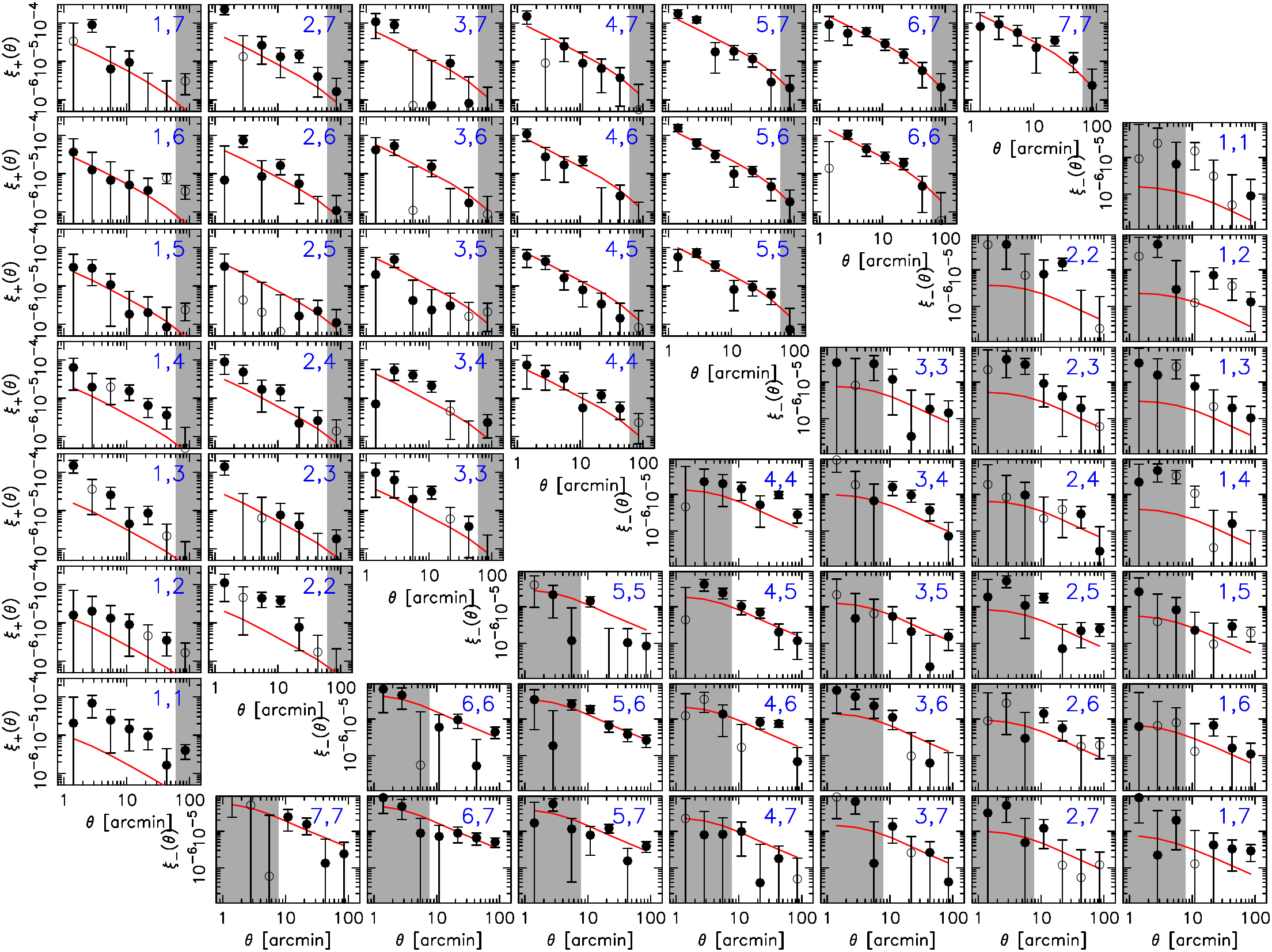}}}
\end{center}
\caption{Measurements of the cosmic shear statistics $\xi_+$ (upper triangle) and $\xi_-$ (lower triangle) against angular scale in arcminutes for all unique pairs of the 7 tomographic source bins, defined in Section~\ref{measlab}. The error bars are determined using the mock catalogues described in Section~\ref{covlab}.
The grey regions correspond to angular scales that were removed from the cosmology analysis, 
due to low signal-to-noise or covariance under-estimation (discussed in Section~\ref{measlab}). 
Open circles denote negative points. Fiducial theory lines have been included in red (solid) for comparison.
}
\label{figshear}
\end{figure*}

For each unique pair of tomographic bins, we measured the cosmic shear
statistics $(\xi_+, \xi_-)$ in each of the 4 regions using the \athena software (\citealt{2014ascl.soft02026K}).  We use $N_\theta = 7$ equally-spaced logarithmic
bins in the range $1 < \theta < 120$ arcmin. 
Concretely, for each of the 7 tomographic bins, our measurements are evaluated at $[1.41, 2.79, 5.53, 11.0, 21.7, 43.0, 85.2 ]$ arcmins.
This represents a significant increase in the maximum fitted scale of $\approx 53$ arcmins
used by \citealt{Heymans13} (where the earlier reported range of 1.5 to 35 arcmins corresponded to the central bin values), which is enabled by the increased box size of the
N-body simulations now used to determine the data covariance,
described below. We also determined jackknife errors in our
measurements, splitting the data sample into 
jackknife regions defined by each individual MegaCam field. 
We combined the measurements in the different CFHTLenS regions, weighting by the
$N_{\rm pairs}$ value for each bin returned by \athena.  Figure
\ref{figshear} displays the resulting combined $(\xi_+, \xi_-)$
measurements in panels of pairs of tomographic bins.

We arranged the $(\xi_+, \xi_-)$ measurements into a data vector
following the convention of \citet{Heymans13}, such that for each unique pair of tomographic bins,
the $\xi_+$ values are listed with increasing $\theta$, followed by the $\xi_-$ values.
The pairs of tomographic bins $ij$ are then ordered as $(11, 12, ..., 17, 22, 23,
..., 77)$. 
The length of the full data vector is $p = N_\theta N_t (N_t + 1) = 392$ elements, although this is further pruned before cosmological fitting. 
Concretely, we cut our data vector from $p = 392$ elements to $p = 280$ elements by removing angular bins 1 to 3 for $\xi_-$ and the seventh angular bin for $\xi_+$. This is motivated by low signal-to-noise of the removed elements (bins 1 to 3 for $\xi_-$), along with roughly 10\% covariance under-estimation for $\xi_+$ in the seventh bin due to the finite box size for the simulations~\citep{hdw15}.

Lastly, when comparing cosmological constraints from our updated CFHTLenS cosmic shear tomography measurements with cosmic microwave background measurements from the Planck satelite \citep{planck15,planck15like}, we include both CMB temperature and polarization information for Planck on large angular scales, limited to multipoles $\ell \leq 29$ (i.e.~low-$\ell$ TEB likelihood), and restrict ourselves to CMB temperature information on smaller angular scales (via the \plik TT likelihood).
Thus, we conservatively do not include polarization data for the smaller angular scales and we also do not include Planck CMB lensing measurements.

\subsection{Covariance}
\label{covlab}

We determined the covariance of our $(\xi_+, \xi_-)$ measurements
using a set of mock catalogues created from a large suite of N-body
simulations which include a self-consistent computation of
gravitational lensing.  Our covariance methodology follows the
approach of \citet{Heymans13}, with some enhancements described below.

Our starting point is the SLICS (Scinet LIght Cone Simulations) series \citep{hdw15}, which consists of 
500 N-body dark matter
simulations created with the {\small CUBEP$^3$M} code \citep{hd13} using a WMAP9+BAO+SN cosmological parameter set: matter
density $\Omega_{\mathrm m} = 0.2905$, baryon density $\Omega_{\mathrm b} = 0.0473$,
Hubble parameter $h = 0.6898$, spectral index $n_{\mathrm s} = 0.969$ and
normalization $\sigma_8 = 0.826$. 
Although the simulations are evaluated at a fixed cosmology, 
we assume that the cosmology dependence of the resulting covariance matrix has negligible impact on our cosmological parameter constraints following the explicit demonstration of this for a CFHTLenS-like survey in \citet{kilbinger13} (also see \citealt{eifler09}).
The box-size of the simulations is
$L = 505 \, h^{-1}$ Mpc. This is significantly larger than the
simulation set used for modelling the earlier CFHTLenS measurements
[$L = (147, 231) \, h^{-1}$ Mpc], 
significantly reducing the suppression of the large-scale signal and variance caused by the finite box size.
The simulations follow the nonlinear
evolution of $1536^3$ particles inside a $3072^3$ grid cube.

\begin{figure}
\begin{center}
\resizebox{8cm}{!}{\rotatebox{270}{\includegraphics{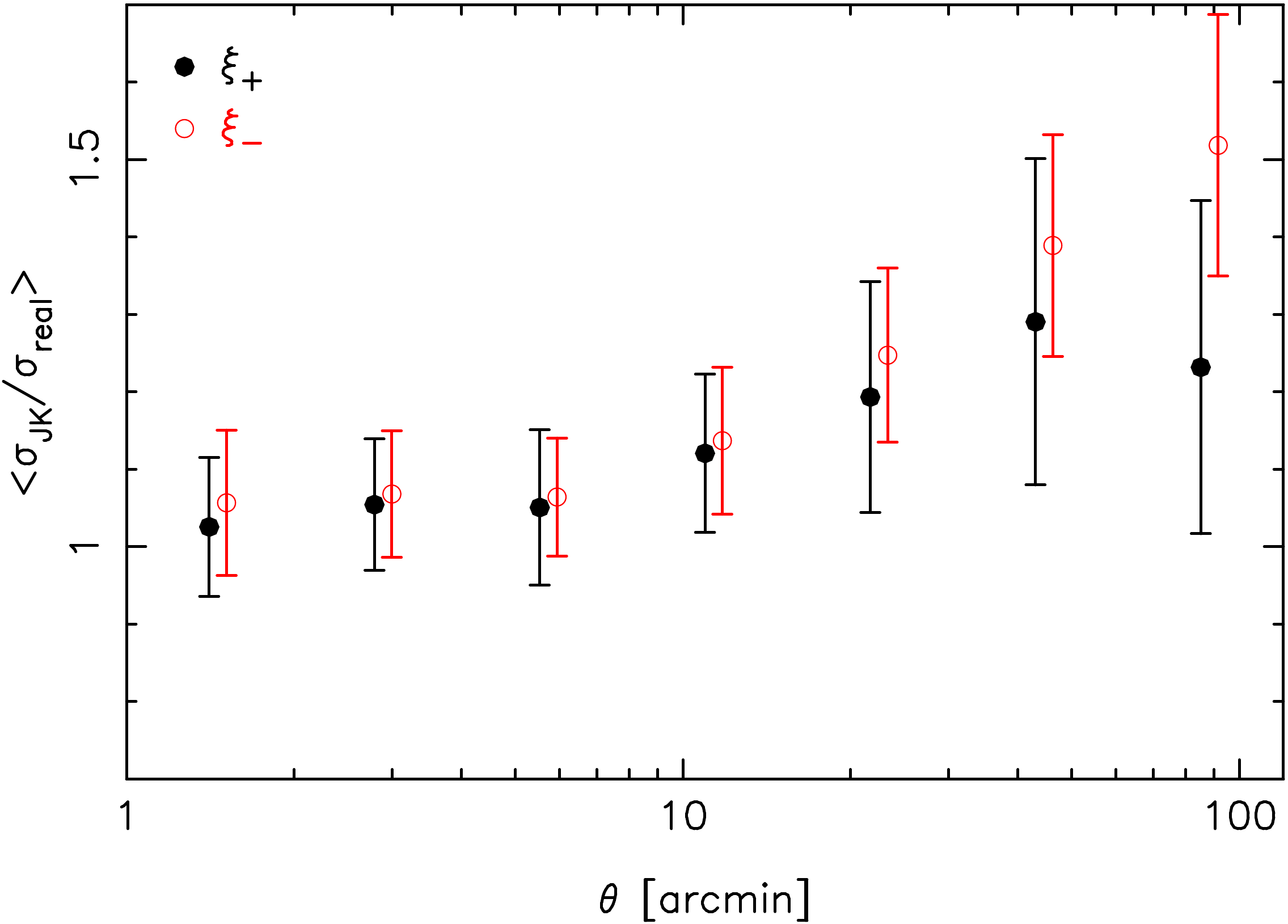}}}
\end{center}
\caption{Ratio of the cosmic shear error determined by jackknife
  sampling using \athena, to that determined from the suite of
  mock catalogues,
  averaged across all tomographic bins as a function of angular scale for $\xi_+$ (black solid circles) and $\xi_-$ (red open circles).}
\label{figerrratio}
\end{figure}

\begin{figure}
\begin{center}
\resizebox{8cm}{!}{\rotatebox{270}{\includegraphics{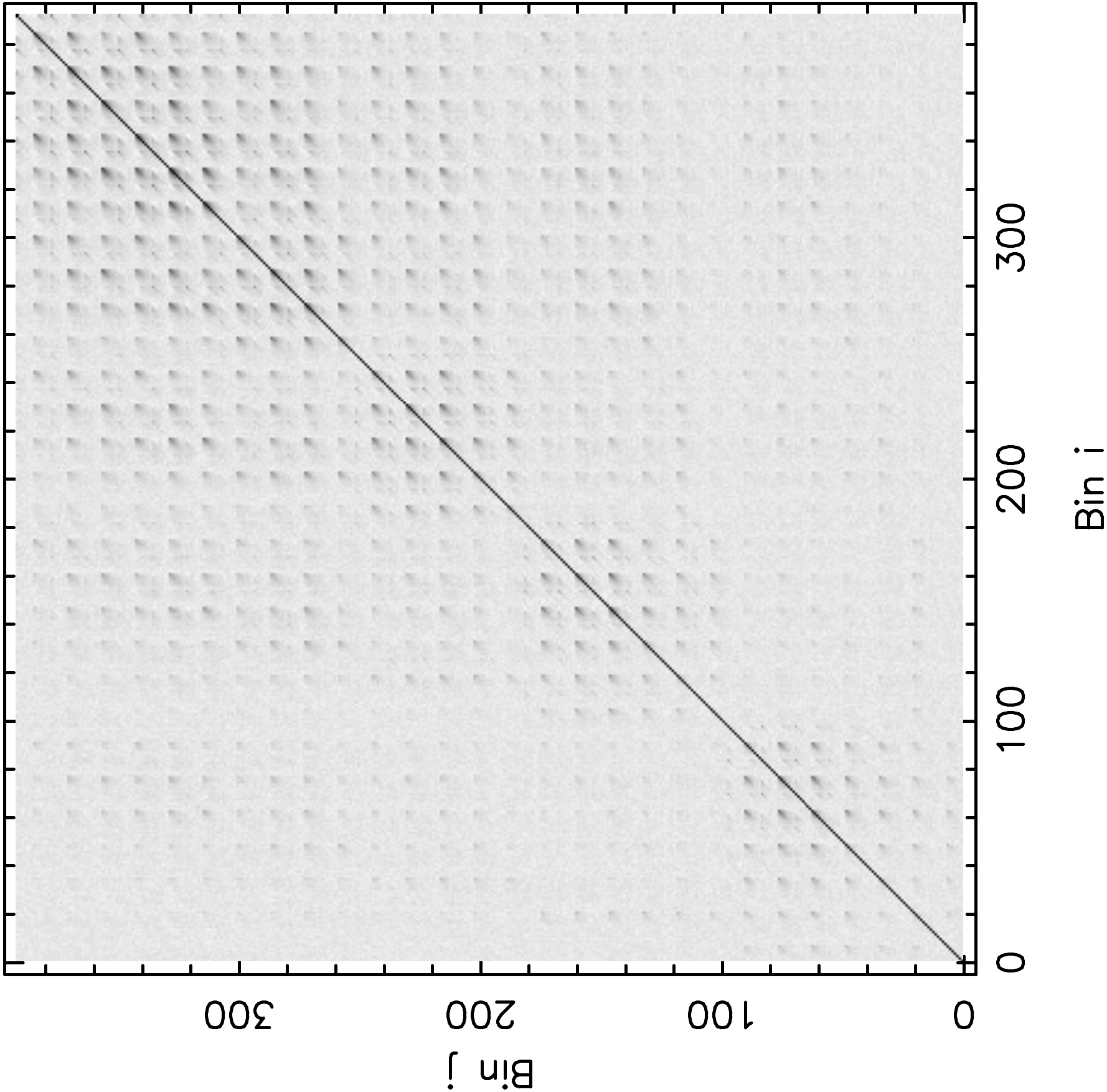}}}
\end{center}
\caption{The correlation coefficient of the covariance matrix of the
  full data vector, plotted using a greyscale where white represents
  $r=0$ and black represents $r=1$.}
\label{figcov}
\end{figure}

For each simulation, the density field is output at 18 redshift
snapshots in the range $0 < z < 3$. The gravitational lensing shear
and convergence are computed at these multiple lens planes using the
Born approximation in the flat-sky approximation, and a survey cone
spanning 60 deg$^2$ is constructed by pasting together these
snapshots.  A dark matter halo finder is also applied to the particle
data at each snapshot, such that self-consistent halo catalogues for
each cone are also produced.

Removing some rare cases of failed simulation outputs (e.g.~due to cooling failure or internode message passing failure from traffic jam in the network), we use 497
independent simulations in our analysis.  We convert these simulation
density and shear fields into mock catalogues for our cosmic shear
covariance using the following process.  We note that although sources
in the dataset have optimal {\it lens}fit weights used in cosmic shear
analysis, we produce mocks in which all sources have uniform weight
$w^s=1$, by building the mocks using the weighted source densities,
weighted redshift distributions and weighted ellipticity variance
measured from the data.

\begin{itemize}
\item In order to gain a sufficient number of realizations that our
  inverse covariance will not be unduly biased by noise in the sample covariance estimator
  \citep{hartlap07}, we {\it split each 60 deg$^2$ simulation box
    into $2 \times 2$ sub-divisions,} producing 1988
  `pseudo-independent' sub-realizations (following \citet{Heymans13}, who used $3
  \times 3$ sub-divisions of a 12.84 deg$^2$ box; each of our
  sub-samples is therefore an order of magnitude larger).

\item We assigned a {\it source redshift distribution} to each survey
  cone with a weighted effective surface density of 10.7 arcmin$^{-2}$
  using the weighted CFHTLenS source redshift probability distribution
  (as described above), Monte-Carlo sampling sources from the density
  field with bias $b_{\rm source} = 1$.  We ensured that sources are
  produced with a continuous distribution in redshift by linearly interpolating the shear
  across the finite redshift width of each snapshot.

\item We assigned the {\it two-component gravitational shears}
  $(\gamma_1, \gamma_2)$ to each source by linearly interpolating the
  shear fields between the values at adjacent snapshot redshifts at
  the source positions.

\item A {\it photometric redshift} $z_B$ was assigned for each mock source using a scattering probability function $p(z_B|z_{\rm sim})$ as a function of its simulation redshift $z_{\rm sim}$.  This scattering function was constructed from the CFHTLenS dataset using a Monte Carlo technique sampling from the full {\tt BPZ} probability distribution of each source, $p_{\rm BPZ}(z)$, together with its measured $z_B$ value.  Specifically, we sampled a redshift $z_{\rm samp}$ from each $p_{\rm BPZ}(z)$ distribution and then binned the values of $(z_B, z_{\rm samp})$, weighting each galaxy by the {\it lens}fit weight, and determining the distribution over $z_B$ for each $z_{\rm samp}$ bin, normalizing such that $\int p(z_B) \, dz_B = 1$.

\item We applied {\it shape noise} to the source catalogues by
  determining the complex noisy shear $e = (\gamma + n)/(1 + n \,
  \gamma^*)$ \citep{seitzschneider97}, where the components of
  observed shear $(e_1, e_2)$ are found as $e = e_1 + i \, e_2$, the
  true shear $\gamma = \gamma_1 + i \, \gamma_2$, and the noise $n =
  n_1 + i \,n_2$.  The noise components $(n_1, n_2)$ are drawn from
  Gaussian distributions with standard deviation $\sigma_e$, which we
  calibrated as a function of $z_B$ using the weighted ellipticity
  variance of the real data:
\begin{equation}
\sigma_e^2 = \sum_i (w^s_i)^2 \, e_i^2 / \sum_i (w^s_i)^2 .
\label{eqsige}
\end{equation}
We find that $\sigma_e$ as a function of $z_B$ ranges between 0.26 and 0.29 with a mean of 0.28.

\item We applied {\it small-scale masks} to each sub-realization using
  the `mosaic masks' provided by the CFHTLenS team.  Given that
  these masks extend beyond the 15 deg$^2$ area of each
  sub-realization, and that we require each sub-realization to possess
  identical masking to avoid introducing spurious noise, we
  consistently applied the same 15 deg$^2$ cut-out from the mask to
  every sub-realization. However, given that the fraction of unmasked
  area varies between the survey regions (owing, for example, to the
  varying stellar density with Galactic latitude) we repeated this
  process using mosaic masks for each of the four survey 
  regions \{W1, W2, W3, W4\}, and derived the final covariance as 
  the area-weighted average of the four determinations.

\end{itemize}

\begin{table*}
\begin{center}
\caption{Exploring the impact of cosmological priors (applicable to Section~\ref{impri} and Fig.~\ref{figprior}). 
The four cases include the same uniform priors on $\{\Omega_c h^2, \Omega_b h^2, \theta_{\rm MC}\}$, and differ in the priors on $\{A_s, n_s, h, k_{\rm pivot}\}$.
Concretely, Cases I and II have wider priors on $\{A_s, n_s, h\}$ than Cases III and IV. The choice of pivot scale further distinguishes Case I from Case II and Case III from Case IV (Planck and WMAP motivated, respectively).
The cosmological parameters in this table are defined as `vanilla' parameters, and $\theta_s$ denotes the angular size of the sound horizon at the redshift of last scattering.
}
\begin{tabular}{p{3.75cm}P{2.85cm}p{2.2cm}p{2.2cm}p{2.2cm}p{1.9cm}}
\toprule
Parameter & Symbol & Prior Case I & Prior Case II & Prior Case III & Prior Case IV\\
\midrule
Cold dark matter density & $\Omega_{c}h^2$ & $0.001 \to 0.99$ & $0.001 \to 0.99$ & $0.001 \to 0.99$ & $0.001 \to 0.99$\\
Baryon density & $\Omega_{b}h^2$ & $0.005 \to 0.1$ & $0.005 \to 0.1$ & $0.005 \to 0.1$ & $0.005 \to 0.1$\\
100 $\times$ approximation to $\theta_s$ & $100 \theta_{\rm MC}$ & $0.5 \to 10$ & $0.5 \to 10$ & $0.5 \to 10$ & $0.5 \to 10$\\
Amplitude of scalar spectrum & $\ln{(10^{10} A_{s})}$ & $1.7 \to 5.0$ & $1.7 \to 5.0$ & $2.3 \to 5.0$ & $2.3 \to 5.0$\\
Scalar spectral index & $n_{s}$ & $0.5 \to 1.5$ & $0.5 \to 1.5$ & $0.7 \to 1.3$ & $0.7 \to 1.3$\\
Dimensionless Hubble constant & $h$ & $0.2 \to 1.4$ & $0.2 \to 1.4$ & $0.4 \to 1.3$ & $0.4 \to 1.3$ \\
Pivot scale $[{\rm{Mpc}}^{-1}]$ & $k_{\rm pivot}$ & 0.05 & 0.002 & 0.05 & 0.002 \\
\bottomrule
\end{tabular}
\label{table:priorcases}
\end{center}
\end{table*}

We hence produced $n_\mu = 1988$ pseudo-independent mock CFHTLenS
shear catalogues, matching the effective source density, underlying
spectroscopic redshift distribution, photo-$z$ scatters,
$z_B$-dependent shape noise, and small-scale masking to the real
dataset.  We divided each mock catalogue by photometric redshift $z_B$
into 7 tomographic redshift bins, and used \athena to measure the
cosmic shear statistics $(\xi_+, \xi_-)$ for the same angular bins as
defined in Section~\ref{measlab}, which we arranged in a data vector $\vec{D}$ (writing the
measurement of bin $i$ in mock $k$ as $D_{ki}$).  We then derived the
data covariance through `area-scaling' as
\begin{equation}
{\rm Cov}(i,j) = \frac{A_{\rm mock}}{A_{\rm eff} \, (n_\mu-1)}
\sum_{k=1}^{n_\mu} \left( D_{ki} - \overline{D_i} \right) \left( D_{kj} -
\overline{D_j} \right) ,
\end{equation}
where $\overline{D_i} = \sum_{k=1}^{n_\mu} D_{ki}$, $A_{\rm eff}$ is
the unmasked area of the data determined above, and $A_{\rm mock}$ is
the unmasked area of the sub-realizations.
For mocks that include separations up to a few degrees the error induced by area-scaling, as evaluated using equation~36 in \citet{friedrichdes}, is small compared to other factors.

The jackknife error estimates are computed by dividing the survey into sub-regions defined by the individual 1~deg$^2$ MegaCam pointings (also see~\citealt{friedrichdes}).
The ratio of the error in each bin determined from the suite of mock
catalogues, to the error determined from jackknife re-sampling, is
displayed in Figure \ref{figerrratio}, for the full data vector of 392
values ordered as described above. 
The ratio is close to unity for
small angular scales where jackknife errors are reliable, 
but the jackknife error exceeds the dispersion of the simulations by a factor of more
than 1.5 on the largest scales.
Figure \ref{figcov} displays the full
covariance matrix in the form of a correlation coefficient, 
\begin{equation}
r(i,j) = {\rm Cov}(i,j) / \sqrt{ {\rm Cov}(i,i) \, {\rm Cov}(j,j) } .
\end{equation}
As in \citet{Heymans13}, we further obtain an unbiased estimate of the inverse covariance matrix by implementing the multiplicative correction advocated by \citet{kaufman67} and \citet{hartlap07}, such that
\begin{equation}
{\bf { Cov}^{-1}_{\rm unbiased}} = {{n_{\mu} - p - 2} \over {n_{\mu} - 1}} {\bf Cov}^{-1} ,
\end{equation}
where $n_{\mu}$ is the number of pseudo-independent realizations and $p$ is the number of data points. For 280 elements in our data vector and 1988 pseudo-independent realizations, we obtain a correction of 0.86, while for a reduced data vector of 56 elements (considered in the `max' case defined in Section~\ref{jointsec}), we obtain a correction of 0.97. For both of these cases, our $p/n_{\mu}$ ratios are sufficiently low to avoid over-estimating our Bayesian confidence regions by more than $\approx5\%$ \citep{hartlap07}. While \citet{sh15} extend the analysis of \citet{hartlap07} by relaxing the assumption of a Gaussian likelihood, 
this is mostly visible in the tail of the distribution and does not significantly affect our analysis.

\begin{figure*}
\vspace{-0.8em}
\begin{center}
\resizebox{8.6cm}{!}{{\includegraphics{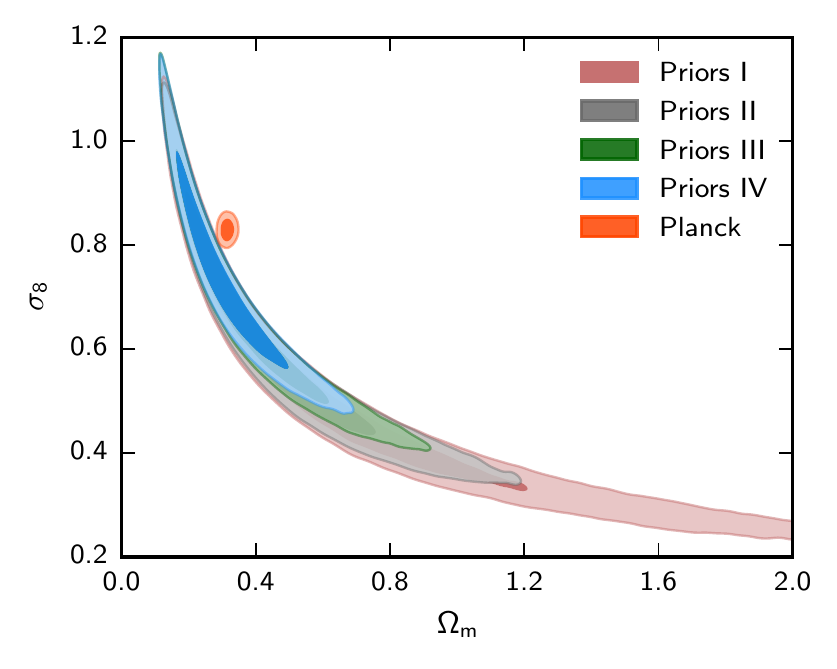}}}
\resizebox{8.8cm}{!}{{\includegraphics{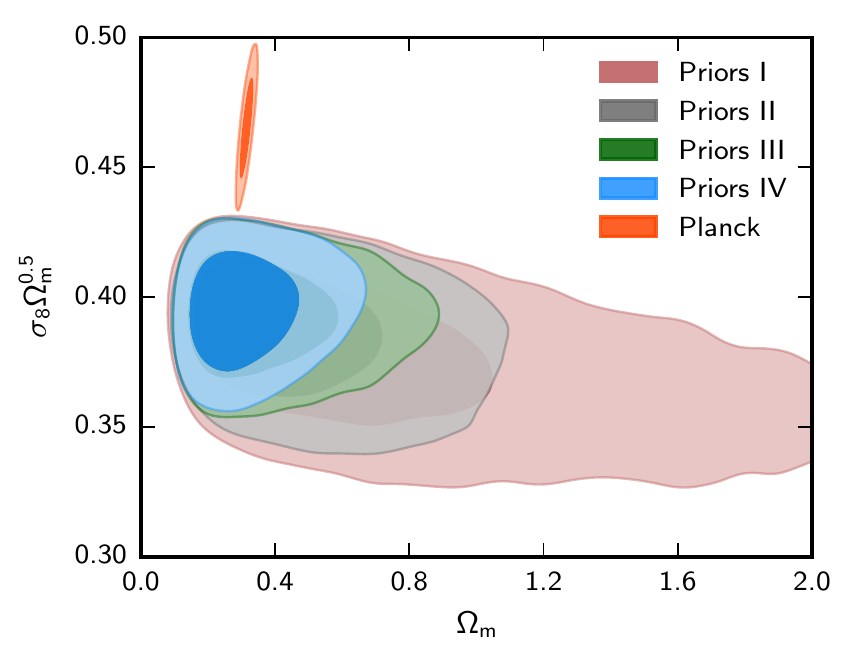}}}
\end{center}
\vspace{-2.2em}
\caption{Marginalized posterior contours in the $\sigma_8 - \Omega_{\mathrm m}$ plane (inner 68\%~CL, outer 95\%~CL) from the updated CFHTLenS cosmic shear tomography measurements with different choices of cosmological priors (purple, grey, green, blue, for Cases I, II, III, IV), 
defined in Table~\ref{table:priorcases}. 
The Planck contour is included for comparison in red (where our Planck dataset is defined in Section~\ref{measlab}).
Right: Same as the left panel, except now showing contours in $\Omega_{\mathrm m}$ against $\sigma_8 \Omega_{\mathrm m}^{0.5}$, orthogonal to the $\sigma_8 - \Omega_{\mathrm m}$ degeneracy direction.
}
\label{figprior}
\end{figure*}

\section{Results}
\label{results}

We now explore the cosmological constraints using the updated CFHTLenS cosmic shear tomography measurements analyzed with the new cosmology fitting pipeline (described in Section~\ref{modtheory}). 
In accordance with \citet{kilbinger13} and \citet{Heymans13}, we mainly illustrate the cosmological constraints in the $\sigma_8 - \Omega_{\mathrm m}$ plane. We begin with a discussion of the constraints when no systematic uncertainties are included, and then methodically include intrinsic alignments, baryonic uncertainties in the nonlinear matter power spectrum, and photometric redshift uncertainties, independently and jointly. We present the main results of these cases associated with the goodness of fit, deviance information criterion, Bayesian evidence, and dataset concordance tests in Tables~\ref{table:pridic}~and~\ref{table:evidence}.

\subsection{Including No Systematic Uncertainties}

\subsubsection{Impact of cosmological priors}
\label{impri}

As a first step, we explore the sensitivity of our weak lensing constraints to the choice of cosmological priors. To this end, we considered four separate cases, listed in Table~\ref{table:priorcases}. All of the cases assume the same broad priors for $\{\Omega_c h^2, \Omega_b h^2, \theta_{\rm MC}\}$, and they differ in the priors for $\{A_s, n_s, h, k_{\rm pivot}\}$. Cases I and II have wider priors on $\{A_s, n_s, h\}$ than Cases III and IV. We moreover allow for either a WMAP-motivated pivot scale $k_{\rm pivot} = 0.002~{\rm{Mpc}}^{-1}$ or a Planck-motivated pivot scale $k_{\rm pivot} = 0.05~{\rm{Mpc}}^{-1}$, 
as a different $k_{\rm pivot}$ translates to different values for $A_s$ and $n_s$ and effectively changes the priors on these parameters.
This choice of pivot scale distinguishes Case I from Case II and Case III from Case IV.

For the four cases considered, in the left panel of Fig~\ref{figprior} we find significant differences in the marginalized posterior contours along the $\sigma_8 - \Omega_{\mathrm m}$ plane. 
However, the four cases show remarkable agreement along the axis perpendicular to the degeneracy direction, such that the $2\sigma$ tension with Planck (reported earlier, e.g.~\citealt{maccrann15}) effectively remains at the same level of significance regardless of the choice of priors. This is further manifested in the right panel of Fig~\ref{figprior}, where we illustrate the constraints on $\sigma_8 \Omega_{\mathrm m}^{0.5}$.
The marginalized posterior contours in the $\sigma_8 - \Omega_{\mathrm m}$ plane shrink when using tighter priors, and also show a sensitivity to the choice of pivot scale.
This implies that current lensing data from CFHTLenS is not sufficiently powerful to constrain the full vanilla parameter space when considering non-informative priors. The only two parameters that are constrained on both ends by the data are $\{\Omega_c h^2, \theta_{\rm MC}\}$, while the other parameters are unconstrained in either one or both directions. 
Moreover, the four cases differ from each other by at most $\Delta\chi^2_{\rm eff} = 2.3$, such that there is no strong statistical preference between the respective best-fit points.

\subsubsection{Choice of cosmological priors}
\label{chopri}

While the contours in Fig~\ref{figprior} could continue to expand by choosing ever more conservative priors, the cosmological priors for the four cases are all objectively conservative, and we do not expect the true values of the parameters to lie outside any of the ranges specified in Table~\ref{table:priorcases}. In order to better understand if there exists a real tension between CFHTLenS and Planck, 
we hereafter adopt our fiducial case, consisting of Case III with
external priors on the Hubble constant and baryon density from Cepheid data and BBN.
Concretely, we uniformly impose the prior $0.61 < h < 0.81$, consistent with \citet{e14} at 99.7\% confidence level (CL), and we uniformly impose the extremely conservative prior $0.013 < \Omega_b h^2 < 0.033$ (allowing for potential systematics and exotic physics), consistent with \citet{burles01,pdg,cyburt15}. 
As these external priors are completely consistent with Planck, any further tension with Planck would therefore derive from CFHTLenS.

\begin{figure}
\vspace{-0.8em}
\begin{center}
\resizebox{8.5cm}{!}{{\includegraphics{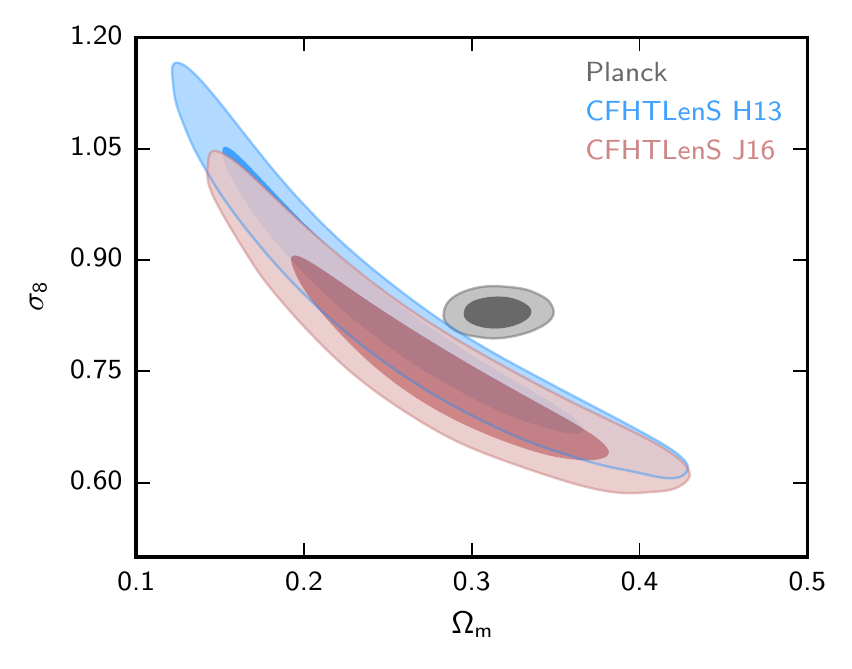}}}
\end{center}
\vspace{-2.2em}
\caption{Marginalized posterior contour in the $\sigma_8 - \Omega_{\mathrm m}$ plane (inner 68\%~CL, outer 95\%~CL) from the updated CFHTLenS cosmic shear tomography measurements (CFHTLenS-J16; in purple),
with fiducial cosmological priors
listed in Table~\ref{table:prisys}. For comparison, including the corresponding contour using the \citet{Heymans13} measurements with our fiducial cosmological priors (CFHTLenS-H13; in blue) and the cosmic microwave background measurements from Planck (in grey).}
\label{figoldnew}
\end{figure}

Given these choices for the priors, we show the resulting contours in the $\sigma_8 - \Omega_{\mathrm m}$ plane in Fig.~\ref{figoldnew}. In addition to using the measurements and covariance described in this paper (denoted `CFHTLenS-J16' with a data vector consisting of 280 elements given 7 tomographic bins), we also show the resulting contour using the original CFHTLenS measurements and covariance from \citet{Heymans13} (denoted `CFHTLenS-H13' with a data vector consisting of 210 elements given 6 tomographic bins). We find that the two analyses agree well, and that there seems to be a marginal increase in the tension with Planck for the new measurements. As for the statistical goodness of the lensing fits, we find $\chi^2_{\rm red} = 1.51$ for the new measurements, as compared to $\chi^2_{\rm red} = 1.19$ for the old measurements.

\begin{table}
\begin{center}
\caption{Priors on the systematic degrees of freedom when considered independently (applicable to Sections~\ref{chopri}~to~\ref{photsec} and Figures~\ref{figoldnew}~to~\ref{figphotoz}).
We also list the fiducial cosmological priors (applicable everywhere from Section~\ref{chopri}).
The `$\to$' sign indicates uniform priors and the `$\pm$' sign indicates Gaussian priors.
When the IA parameters and photo-z bins are fixed, they are set to zero.
Since we do not have external information on the sixth and seventh photo-z bins, when considering an informative photo-z scenario, we keep non-informative priors on these bins.
Please note that we have imposed informative priors on the baryon density and Hubble constant, as described in Section~\ref{chopri}.
The cosmological parameters in this table are defined as `vanilla' parameters, and $\theta_s$ denotes the angular size of the sound horizon at the redshift of last scattering.
}
\begin{tabular}{lc|c}
\toprule
Parameter & Symbol & Prior\\
\midrule
Cold dark matter density & $\Omega_{c}h^2$ & $0.001 \to 0.99$\\
Baryon density & $\Omega_{b}h^2$ & $0.013 \to 0.033$\\
100 $\times$ approximation to $\theta_s$ & $100 \theta_{\rm MC}$ & $0.5 \to 10$\\
Amplitude of scalar spectrum & $\ln{(10^{10} A_{s})}$ & $2.3 \to 5.0$\\
Scalar spectral index & $n_{s}$ & $0.7 \to 1.3$\\
Dimensionless Hubble constant & $h$ & $0.61 \to 0.81$ \\
Pivot scale $[{\rm{Mpc}}^{-1}]$ & $k_{\rm pivot}$ & 0.05 \\
\midrule
IA amplitude & $A$ & $-50 \to 50$\\
{\it~~~--~informative case} & $$ &  $-6 \to 6$\\
IA redshift dependence & $\eta$ & $-50 \to 50$\\
{\it~~~--~informative case} & $$ &  $0$\\
IA luminosity dependence & $\beta$ & $-50 \to 50$\\
{\it~~~--~informative case} & $$ &  $1.13 \pm 0.25$\\
\midrule
\hmcode feedback amplitude & $\log{B}$ & $0 \to 2$\\
{\it~~~--~informative case} & $$ &  $0.3 \to 0.6$\\
{\it~~~--~when fixed} & $$ &  $0.496$\\
\midrule
Photo-z bin 1 & ${\Delta}z_1$ & $-0.1 \to 0.1$\\
{\it~~~--~informative case} & $$ &  $-0.045 \pm 0.013$\\
Photo-z bin 2 & ${\Delta}z_2$ & $-0.1 \to 0.1$\\
{\it~~~--~informative case} & $$ &  $-0.014 \pm 0.010$\\
Photo-z bin 3 & ${\Delta}z_3$ & $-0.1 \to 0.1$\\
{\it~~~--~informative case} & $$ &  $0.008 \pm 0.008$\\
Photo-z bin 4 & ${\Delta}z_4$ & $-0.1 \to 0.1$\\
{\it~~~--~informative case} & $$ &  $0.042 \pm 0.017$\\
Photo-z bin 5 & ${\Delta}z_5$ & $-0.1 \to 0.1$\\
{\it~~~--~informative case} & $$ &  $0.042 \pm 0.034$\\
Photo-z bin 6 & ${\Delta}z_6$ & $-0.1 \to 0.1$\\
Photo-z bin 7 & ${\Delta}z_7$ & $-0.1 \to 0.1$\\
\bottomrule
\end{tabular}
\label{table:prisys}
\end{center}
\end{table}

\begin{table}
\begin{center}
\caption{Exploring changes in $\chi^2_{\rm eff}$ and DIC for different choices of systematic uncertainties given fiducial cosmological priors.
The reference vanilla model without systematic uncertainties gives $\chi^2_{\rm eff} = 414.6$ and $\rm{DIC} = 421.7$ when using \halofit, and $\chi^2_{\rm eff} = 416.4$ and $\rm{DIC} = 423.3$ when using \hmcode (with fixed $\log B = 0.496$). Since the size of the data vector for the `max' case is significantly smaller than the size of the fiducial data vector (where the max case keeps only `large' angular scales and is defined in Section~\ref{jointsec}), we calculate the difference in $\chi^2_{\rm eff}$ and DIC with respect to the measurements used for the `max' case but without systematic uncertainties. 
For this reduced data vector, $\chi^2_{\rm eff} = 86.8$ and $\rm{DIC} = 92.0$, considering \hmcode with $\log B = 0.496$.
}
\begin{tabular}{p{2.4cm}>{\raggedleft}p{2.6cm}>{\raggedleft\arraybackslash}p{2.0cm}}
\toprule
Model & $\Delta\chi^2_{\rm eff}$ & $\Delta{\rm DIC}$\\
\midrule
vanilla + $A$ & $-5.8$ & $-4.6$\\
{\it~~~--~informative case} & $-5.7$ & $-4.9$\\
vanilla + $\{A,\eta,\beta\}$ & $-21$ & $12$\\
{\it~~~--~informative case} & $-0.72$ & $2.4$\\
vanilla + $B$ & $-1.9$ & $-0.64$\\
{\it~~~--~informative case} & $-0.22$ & $1.4$\\
vanilla + 7 photo-z & $-6.0$ & $0.97$\\
{\it~~~--~informative case} & $1.5$ & $11$\\
vanilla + min case & $-10$ & $1.6$\\
vanilla + mid case & $-1.1$ & $12$\\
vanilla + max case & $-25$ & $19$ \\
\bottomrule
\end{tabular}
\label{table:pridic}
\end{center}
\end{table}

The reduction in the `goodness of fit' between the two analyses derives from two changes in the analysis. The first change is the use of a new suite of N-body simulations to determine the covariance matrix. In \citet{Heymans13}, the field-of-view of the 184 simulations used was only 12.84~deg$^2$. In order to gain enough mock realizations to accurately invert the covariance matrix, they split the simulations into $3 \times 3$ sub-realizations such that each sub-realization was close in size to the $\approx53$ arcmins maximum scale measured for the lensing statistics. Pairs on those scales were therefore `missing' due to edge effects and as a result the error on large scales was overestimated. In our analysis, the field-of-view of the 497 simulations used is 60~deg$^2$ and we can therefore measure the large-scale simulated covariance accurately. As the CFHTLenS data is a poorer fit to the model on large scales, the reduction in errors on large scales results in an increased $\chi^2_{\rm red}$.

While our new covariance analysis is certainly an improvement on \citet{Heymans13}, it also does not include super-sample variance terms \citep{thu13}. These super-sampling variance errors contribute to all angular scales and are missing from our calculation as very large-scale modes in the density field are not simulated in the finite box of the N-body simulations. 
However, from the good agreement between the jackknife and simulated errors in Fig.~\ref{figerrratio}, we can conclude that these super-sample terms are not significant on small scales where the majority of the cosmological information is accessed. On large scales, including super-sample terms is likely to improve the goodness of fit of the data, an analysis that we will pursue in future work.

The second change in our analysis is the use of angular scales larger than the 50 arcmin limit of \citet{Heymans13}, introduced owing to the limitation of their simulations.~\citet{asgari16} have recently presented an optimal E/B mode decomposition analysis of CFHTLenS using the COSEBIs statistic (\citealt{sek10, asgari12}). This analysis reveals significant B-modes on large angular scales ($\theta > 40$~arcmins) that do not derive from gravitational lensing, which exhibits a pure E-mode signal. These B-modes are further enhanced when the data is analyzed in tomographic bins. 

\citet{asgari16} also present a compressed-COSEBIs analysis of CFHTLenS (formalism introduced in \citealt{asgari15}), where the COSEBIs are optimally combined to extract cosmological information. In this compressed analysis the recovered B-modes are consistent with zero (except for the blue galaxy case in the 40 to 100 arcmins range with six tomographic bins). If we assume that the systematics that introduce B-modes into the data contribute equally to the E- and B-modes, we can conclude that these systematics will impact on the goodness of fit of the E-mode, particularly on large scales where the B-modes are found to be at their strongest. However, as the compressed cosmological parameter analysis results in a zero B-mode, in particular when the full galaxy sample is considered, these B-modes are not degenerate with cosmological parameters and are therefore fairly benign in the cosmological analysis that follows, particularly when we allow for uncertainty in the three astrophysical sources of systematics that we focus on in this paper. We will investigate the origin of these B-modes further in future work.

We now proceed to exploring the impact of three distinct systematic uncertainties on our results:~intrinsic galaxy alignments, baryonic uncertainties in the nonlinear matter power spectrum, and photometric redshift uncertainties. 

\begin{figure*}
\begin{center}
\resizebox{8.6cm}{!}{{\includegraphics{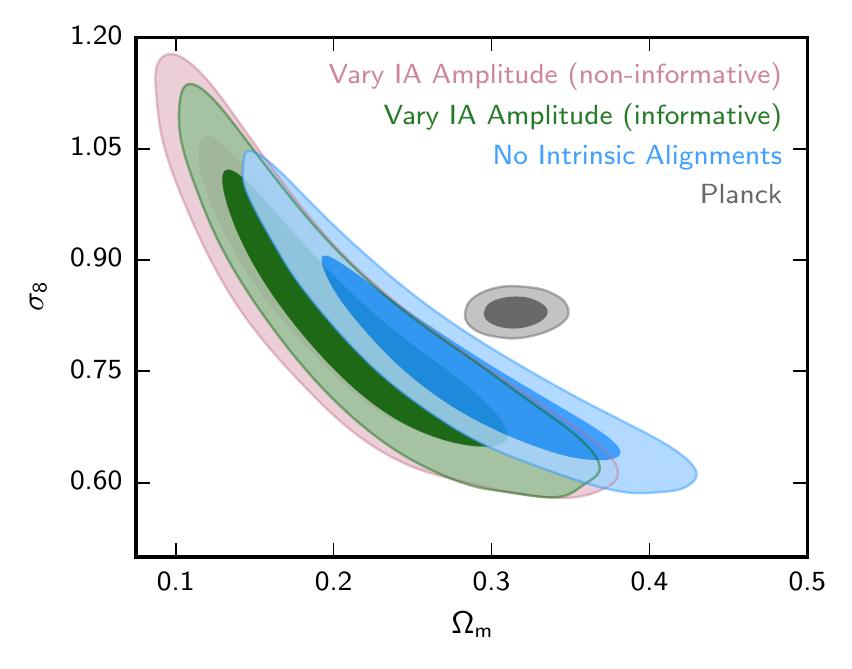}}}
\resizebox{8.6cm}{!}{{\includegraphics{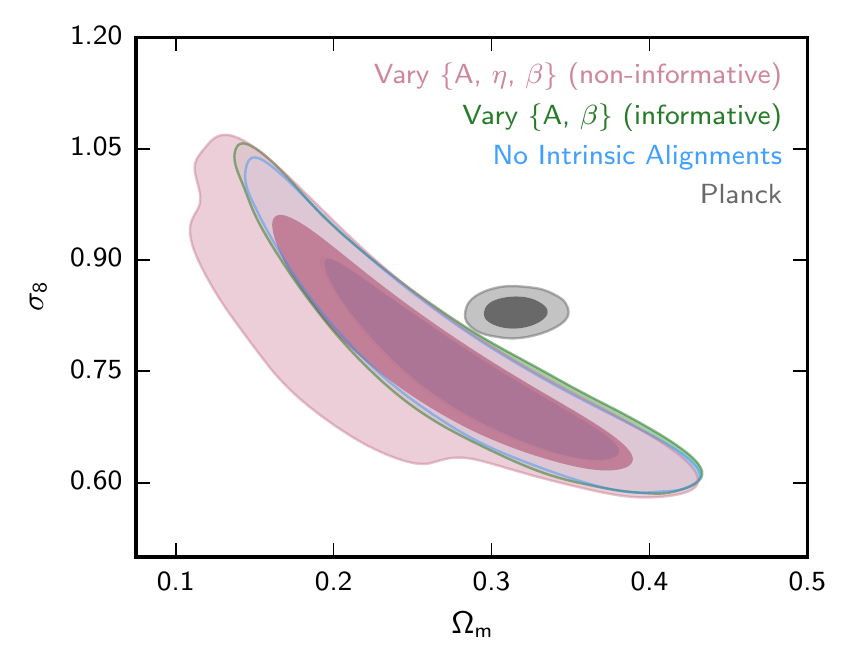}}}
\end{center}
\vspace{-2.3em}
\caption{Left: Marginalized posterior contours in the $\sigma_8 - \Omega_{\mathrm m}$ plane (inner 68\%~CL, outer 95\%~CL) from the updated CFHTLenS cosmic shear tomography measurements, considering fiducial cosmological priors, where the intrinsic alignment amplitude $A$ is allowed to vary with both informative and non-informative priors on the amplitude (in green and purple, respectively). We moreover include the fiducial (no-systematics) cosmic shear and Planck CMB contours for comparison (in blue and grey, respectively). The fiducial cosmological and IA priors are listed in Table~\ref{table:prisys}.
Right: Marginalized posterior contours where all three intrinsic alignment parameters $\{A, \eta, \beta\}$, 
encapsulating the amplitude, redshift dependence, and luminosity dependence of the intrinsic alignments, respectively, are allowed to vary jointly (with both informative and non-informative priors on the three parameters, in green and purple, respectively). Fiducial cosmic shear and Planck CMB contours included for comparison (in blue and grey, respectively).
}
\label{figia}
\end{figure*}

\subsection{Including Intrinsic Galaxy Alignments}
\label{iasec}

We begin by including the three systematic uncertainties independently, before accounting for them jointly. The first of these systematic uncertainties comes from the intrinsic alignments of galaxies. 
We consider two separate scenarios, one in which we only allow for a variation of the amplitude $A$ (defined in equation~\ref{eqn:fz}), and a second scenario in which we also allow for a possible redshift and luminosity dependence of the intrinsic alignment signal via the two parameters $\eta$ and $\beta$, respectively. For each of the two scenarios, we consider both informative and non-informative priors. We specify these priors in Table~\ref{table:prisys}. 

More specifically, when varying the vanilla parameters along with $A$, we let the amplitude vary uniformly between \{-50, 50\} for the non-informative case, and uniformly between \{-6, 6\} for the informative case. When varying the vanilla parameters with $\{A, \eta, \beta\}$, we let each of the intrinsic alignment parameters vary uniformly between \{-50, 50\} for the non-informative case. For the informative case, we let $A$ vary uniformly between \{-6, 6\}, we fix $\eta = 0$, and we impose $\beta = 1.13 \pm 0.25$ as a Gaussian prior (motivated by~\citealt{Joachimi11}).

While our non-informative priors are reasonably wide, our informative priors are driven by the fact that our sample is dominated by blue galaxies, which are known to be less sensitive to intrinsic alignment effects. We have therefore taken the tightest available luminosity and redshift dependent constraints determined from red galaxies as the `worst-case-scenario' for the luminosity and redshift dependence of the dominant blue galaxies in the sample (from \citealt{Joachimi11}, consistent with \citealt{singh15}), while also encompassing the luminosity and redshift dependence in the red sample. For the informative case, we set $\eta = 0$ given the lack of evidence for redshift evolution in \citet{Joachimi11} and \citet{singh15}. 
We further allow for negative values of the intrinsic alignment amplitude as the best-fit model for a mixed population could result in such values, as discussed in \citet{Heymans13}.

In Fig.~\ref{figia}, we show the marginalized posterior contours in the $\sigma_8 - \Omega_{\mathrm m}$ plane for different IA models. When varying the vanilla parameters with the intrinsic alignment amplitude (i.e.~without luminosity or redshift dependence), the contours slightly degrade due to the extra degree of freedom despite the additional cosmological information contained in the II and GI terms (described in Section~\ref{iatheory}). The contours moreover shift towards larger values of $\sigma_8$ and smaller values of the matter density (evident from equation~\ref{eqn:fz}), increasing the tension with Planck. The degradation and shift in the contours are in agreement with \citet{Heymans13}, who pointed out that $\sigma_8$ is driven towards larger values by negative best-fit estimates of the intrinsic alignment amplitude. 

The observed behavior applies to both the non-informative and informative cases, as the non-informative constraint on the intrinsic alignment amplitude is $A = -3.6 \pm 1.6$ (corresponding to the mean of the posterior distribution along with the symmetric 68\% confidence interval about the mean), which implies that the informative case will give similar constraints.
As seen in Table~\ref{table:pridic}, the informative and non-informative cases only differ from each other by $\Delta\chi^2_{\rm eff} = 0.1$. They further differ from the~$A=0$~scenario by $\Delta\chi^2_{\rm eff} = -5.8$ for the non-informative case and $\Delta\chi^2_{\rm eff} = -5.7$ for the informative case. 
The penalty due to the increased Bayesian complexity gives $\Delta{\rm DIC} = -4.6$ for the non-informative case and $\Delta{\rm DIC} = -4.9$ for the informative case. 

\begin{figure}
\vspace{-0.7em}
\begin{center}
\resizebox{8.4cm}{!}{{\includegraphics{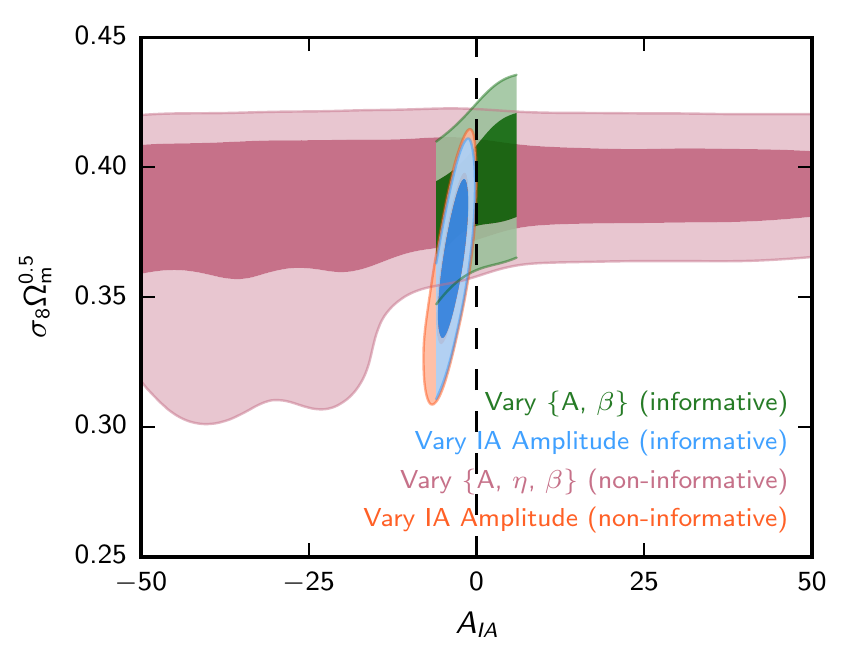}}}
\end{center}
\vspace{-2.2em}
\caption{Marginalized posterior contours in the plane given by $\sigma_8 \Omega_{\mathrm m}^{0.5}$ and intrinsic alignment amplitude $A$ (inner 68\%~CL, outer 95\%~CL) from the updated CFHTLenS cosmic shear tomography measurements, considering both informative and non-informative priors on the intrinsic alignment parameters $\{A, \eta, \beta\}$, encapsulating the amplitude, redshift dependence, and luminosity dependence of the intrinsic alignments, respectively. The fiducial cosmological and IA priors are listed in Table~\ref{table:prisys}. 
}
\label{figampia}
\end{figure}

Thus, there seems to be substantial preference in favor of a nonzero and negative intrinsic alignment amplitude (as also found in Fig.~\ref{figampia}). 
While the intrinsic alignment amplitude for a single sample must be positive, for a mixed sample the best-fit model can be negative, as described in \citet{Heymans13}. Thus, this could be a sign that the model is a good fit to red galaxies, while the majority of our sample is blue. Alternatively, the negative intrinsic alignment amplitude could imply that the IA model we use is too simplistic, or the result of unaccounted systematics. For instance, photometric redshift errors could also mimic IA-like behavior, so this could be a sign that the low-redshift source distributions are inaccurate.

\begin{figure*}
\begin{center}
\resizebox{8.6cm}{!}{{\includegraphics{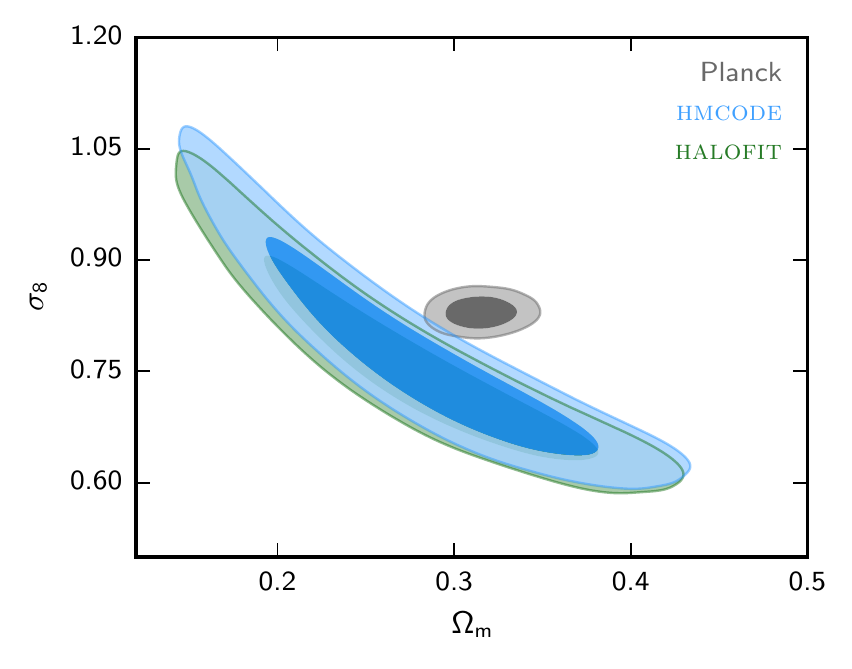}}}
\resizebox{8.6cm}{!}{{\includegraphics{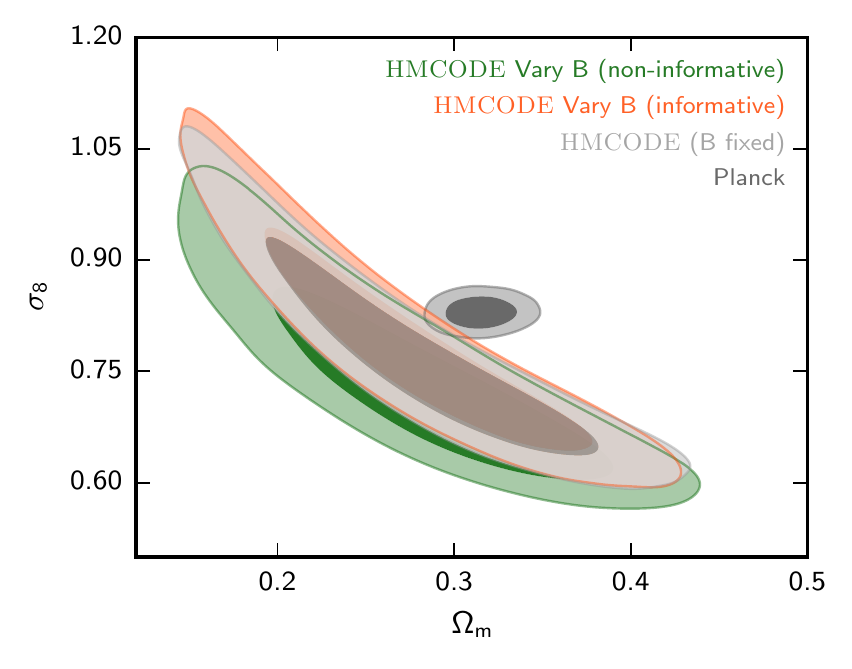}}}
\end{center}
\vspace{-2.3em}
\caption{Left: Marginalized posterior contours in the $\sigma_8 - \Omega_{\mathrm m}$ plane (inner 68\%~CL, outer 95\%~CL) from the updated CFHTLenS cosmic shear tomography measurements when using either \hmcode or \halofit for the nonlinear matter power spectrum (in blue and green, respectively), where the \hmcode feedback amplitude $\log B$ is fixed to its fiducial DM-only value of 0.496. The Planck contour is included for comparison (in grey), and the cosmological priors are listed in Table~\ref{table:prisys}.
Right: Marginalized posterior contours where the \hmcode feedback amplitude is allowed to vary (with both informative and non-informative priors on the amplitude, listed in Table~\ref{table:prisys}, in red and green, respectively). The fiducial cosmic shear contour (where $\log B = 0.496$) and the Planck CMB contour are included for comparison (in brown and grey, respectively).
}
\label{figmead}
\end{figure*}

\begin{figure}
\vspace{-0.8em}
\begin{center}
\resizebox{8.4cm}{!}{{\includegraphics{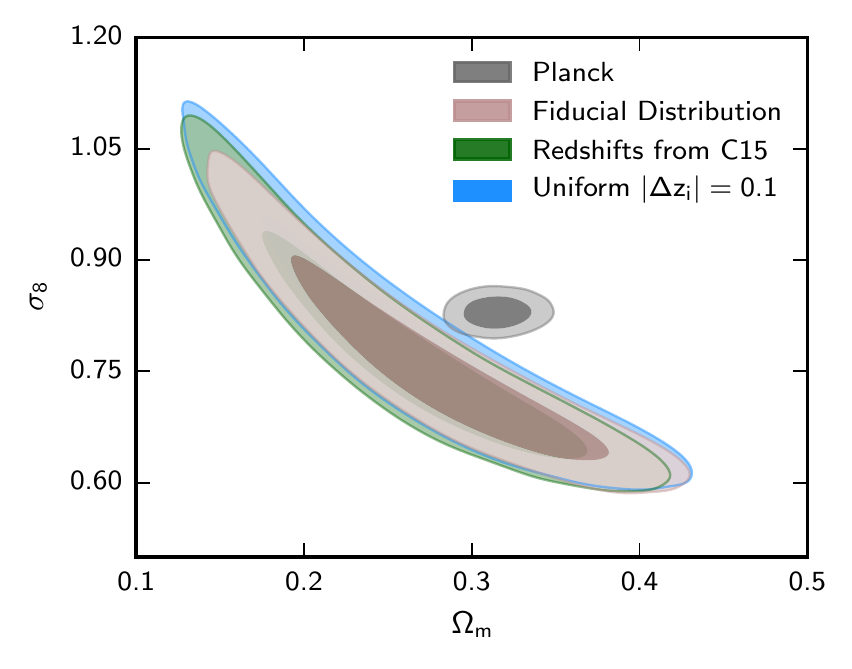}}}
\end{center}
\vspace{-2.2em}
\caption{Marginalized posterior contours in the $\sigma_8 - \Omega_{\mathrm m}$ plane (inner 68\%~CL, outer 95\%~CL) from the updated CFHTLenS cosmic shear tomography measurements for different treatments of the photometric redshift uncertainties. 
The contour where the fiducial redshift distribution is used is given in brown, the contour where the redshift distribution is perturbed according to the results from source-lens cross-correlations in \citet{choi15} is given in green, and the contour where the redshift distribution is perturbed with uniform priors of $\left|{\Delta}z_i\right| = 0.1$ in each tomographic bin is given in blue.
The Planck contour is included for comparison in grey, while the fiducial cosmological and photometric redshift priors are listed in Table~\ref{table:prisys}.
}
\label{figphotoz}
\end{figure}

In Fig.~\ref{figia}, we further show the cosmological constraints as we let all three intrinsic alignment parameters vary simultaneously. We note that this 3-parameter model has not been fit to large-scale cosmic shear data before, for example \citet{Heymans13} only considered varying the IA amplitude and \citet{dessv} considered a 2-parameter model with a varying amplitude and redshift dependence. 
When imposing informative priors on $\{A, \eta, \beta\}$,
the additional degree of freedom from the luminosity dependence causes the contour in the $\sigma_8 - \Omega_{\mathrm m}$ plane to effectively transform back to the original contour given by the $A=0$ scenario. This is because our prior on $\beta$ decreases the strength of the intrinsic alignment II and GI terms in the lensing calculation, even though $A$ is unbounded within the region given by the prior. 
Hence, the cosmological constraints with a large value of $\beta$ mimic the constraints for the scenario with no intrinsic alignments. There is therefore less tension with Planck for this model than the 1-parameter intrinsic alignment model that we first considered. Here, we find $\Delta\chi^2_{\rm eff} = -0.72$, while $\Delta{\rm DIC} = 2.4$ shows a weak preference against the extended intrinsic alignment model. 

When considering non-informative priors on $\{A, \eta, \beta\}$, we find an enlarged contour 
towards smaller values of the matter density (where the enlarged region is consistent with extremely negative values of the IA amplitude as seen in Fig.~\ref{figampia}). 
While the intrinsic alignment amplitude is completely unconstrained, 
we find 1-sided bounds on $-50 < \eta < 8.0$ and $0.54 < \beta < 50$ at 95\% CL (where $\beta > 0$ even at 99.7\% CL). In other words, $\eta$ and $\beta$ are taking on values that exclude potentially enormous intrinsic alignment signals, and are in fact consistent with a negligible signal. 
The range for $\eta$ is moreover consistent with no redshift evolution, while the range for $\beta$ shows a weak preference for a luminosity dependence of the intrinsic alignments.
Interestingly, we find $\Delta\chi^2_{\rm eff} = -21$ for this 3-parameter intrinsic alignment model, while the increased Bayesian complexity of the model is severely penalized in $\Delta{\rm DIC} = 12$. This demonstrates the extreme usefulness of the information criterion, as it determines the 3-parameter intrinsic alignment model to be less preferred than the reference model without intrinsic alignments.

\begin{table*}
\begin{center}
\caption{Exploring changes in systematic priors for three joint scenarios (applicable to Section~\ref{jointsec} and Fig.~\ref{figjoint}).
The `min' case corresponds to the most optimistic scenario for the priors, and the `max' case corresponds to the most conservative scenario, while the `mid' case lies between these two scenarios. In our language, keeping `all' angular scales implies the data vector consists of 280 elements, while keeping `large' angular scales implies the data vector consists of 56 elements, as discussed in Section~\ref{jointsec}.
The `$\to$' sign indicates uniform priors and the `$\pm$' sign indicates Gaussian priors.
Moreover, the priors on the underlying cosmology are the same as in Table~\ref{table:prisys}.
}
\begin{tabular}{p{3.42cm}P{3.72cm}P{2.94cm}P{2.94cm}P{2.44cm}}
\toprule
Parameter & Symbol & Min Case & Mid Case & Max Case\\
\midrule
IA amplitude & $A$ & $-6 \to 6$ & $-6 \to 6$ & $-50 \to 50$\\
IA luminosity dependence & $\beta$ & $0$ & $1.13 \pm 0.25$ & $-50 \to 50$\\
IA redshift dependence & $\eta$ & $0$ & $0$ & $-50 \to 50$\\
\hmcode feedback amplitude & $\log{B}$ & $0.3 \to 0.6$ & $0.3 \to 0.6$ & $0 \to 2$\\
Photo-z bin 1 & ${\Delta}z_1$ & $-0.045 \pm 0.013$ & $-0.045 \pm 0.050$ & $-0.1 \to 0.1$\\
Photo-z bin 2 & ${\Delta}z_2$ & $-0.014 \pm 0.010$ & $-0.014 \pm 0.050$ & $-0.1 \to 0.1$\\
Photo-z bin 3 & ${\Delta}z_3$ & $0.008 \pm 0.008$ & $0.008 \pm 0.050$ & $-0.1 \to 0.1$\\
Photo-z bin 4 & ${\Delta}z_4$ & $0.042 \pm 0.017$ & $0.042 \pm 0.050$ & $-0.1 \to 0.1$\\
Photo-z bin 5 & ${\Delta}z_5$ & $0.042 \pm 0.034$ & $0.042 \pm 0.050$  & $-0.1 \to 0.1$\\
Photo-z bin 6 & ${\Delta}z_6$ & $-0.1 \to 0.1$ & $-0.1 \to 0.1$ & $-0.1 \to 0.1$\\
Photo-z bin 7 & ${\Delta}z_7$ & $-0.1 \to 0.1$ & $-0.1 \to 0.1$ & $-0.1 \to 0.1$\\
Angular scales & $\theta$ & $\rm{All}$ & $\rm{All}$ & $\rm{Large}$\\
\bottomrule
\end{tabular}
\label{table:priorjoint}
\end{center}
\end{table*}

\subsection{Including Baryonic Uncertainties in the Nonlinear Matter Power Spectrum}
\label{barsec}

We now proceed to another important systematic coming from baryonic uncertainties in the nonlinear matter power spectrum. We account for the baryonic effects by varying the \hmcode feedback amplitude $B$, described in Section~\ref{barytheory}. For the scenario with only dark matter, we fix $\log B = 0.496$ as advocated in \citet{Mead15}. 

In Fig.~\ref{figmead}, we first show the marginalized posterior contours in the $\sigma_8 - \Omega_{\mathrm m}$ plane corresponding to the use of either \halofit or \hmcode for the nonlinear extension to the matter power spectrum, considering no baryonic effects on nonlinear scales (such that $\log B$ is fixed to its fiducial value of 0.496 for \hmcode). As one would expect from \citet{Mead15}, which is in excellent agreement with \citet{Takahashi12} for cosmologies where no baryonic effects are included, the contours agree remarkably well for this DM-only scenario. This implies the two matter power spectrum prescriptions can be used interchangeably when baryonic effects are not 
included on nonlinear scales. 
Since \halofit is faster than \hmcode (as discussed in Section~\ref{modtheory}), this has allowed us to comfortably use \halofit for our runs where \hmcode is not directly needed (i.e. when nonlinear baryonic effects are not included). As for the relative fits for the DM-only scenario, using either \hmcode or \halofit, we find a difference of $\Delta\chi^2_{\rm eff} = 1.8$ and $\Delta{\rm DIC} = 1.5$ between the models, such that they are close to equally preferred (with the weak preference in favor of \halofit).

In Fig.~\ref{figmead}, we further show the constraints when allowing the amplitude $B$ in \hmcode to vary freely, considering both informative and non-informative priors (listed in Table~\ref{table:prisys}). We take the informative prior to uniformly cover the range $0.3 < \log B < 0.6$, which effectively corresponds to the range given by the best-fit values of the \{DMONLY, REF, DBLIM, AGN\} cases in the OWL simulations~\citep{Schaye10,Daalen11}, as demonstrated by \citet{Mead15}. 
This informative case gives a contour that marginally prefers larger values of $\sigma_8$, while $\Delta\chi^2_{\rm eff} = -0.22$ and $\Delta{\rm DIC} = 1.4$ (as compared to the case where $\log B$ is fixed to 0.496), suggesting there is no strong preference for or against the additional degree of freedom.
In other words, the constraining power of the weak lensing dataset from CFHTLenS 
seems to be inadequate to distinguish the DM-only model for the nonlinear matter power spectrum from one of the models of the OWL simulations.

For the non-informative case, the \hmcode feedback amplitude is constrained by the data such that $0.37 < \log B < 1.1$ at 95\% CL. The resulting contour is both expanded and shifted to smaller values of $\sigma_8$ (in the plane with $\Omega_{\mathrm m}$). This downward shift in $\sigma_8$ is caused by unnaturally large values of $B$ that are allowed by the data, despite the baryonic OWL simulation models preferring values of $B$ below the fiducial value (as $B$ and $\sigma_8$ are anti-correlated). Thus, as compared to the fiducial model, allowing for an extra degree of freedom in the nonlinear matter power spectrum does not seem to alleviate the tension between CFHTLenS and Planck. Moreover, $\Delta\chi^2_{\rm eff} = -1.9$ and $\Delta{\rm DIC} = -0.64$, which again implies the data does not strongly prefer $B$ to stray away from its fiducial DM-only value.

\begin{figure*}
\begin{center}
\resizebox{8.63cm}{!}{{\includegraphics{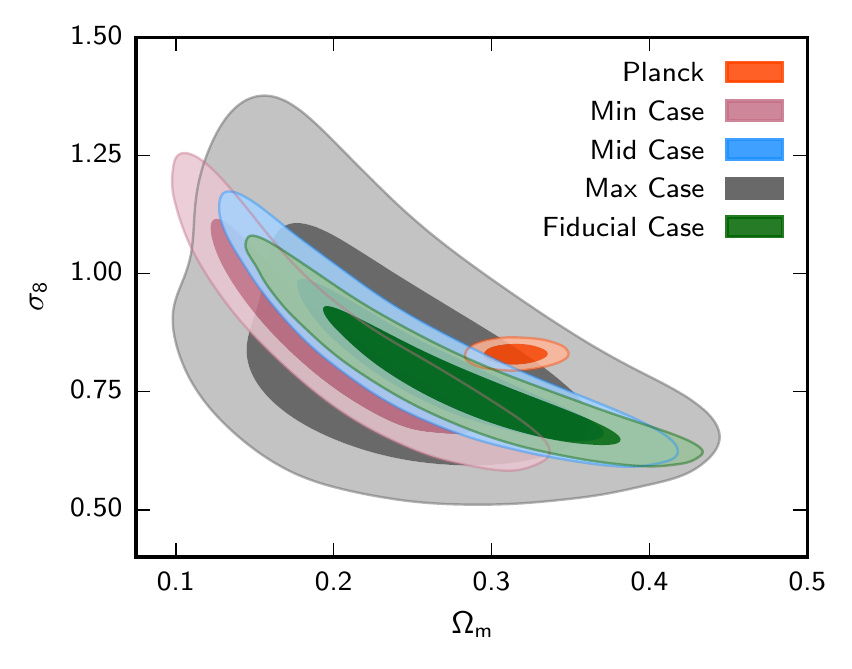}}}
\resizebox{8.38cm}{!}{{\includegraphics{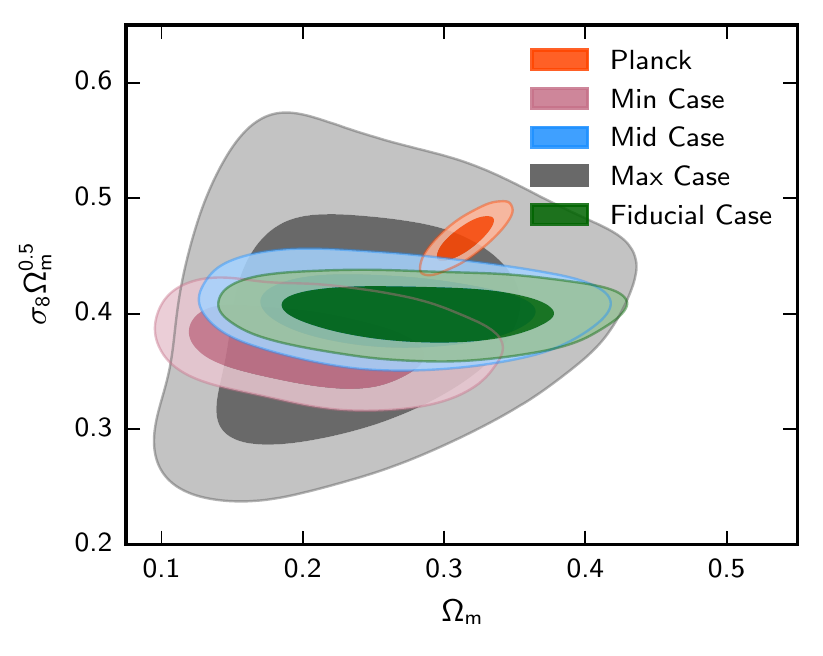}}}
\end{center}
\vspace{-2.2em}
\caption{Left: Marginalized posterior contours in the $\sigma_8 - \Omega_{\mathrm m}$ plane (inner 68\%~CL, outer 95\%~CL) from the updated CFHTLenS cosmic shear tomography measurements for the joint analysis of the systematic uncertainties, where the priors on the cosmological and systematic degrees of freedom are listed in Tables~\ref{table:prisys}~and~\ref{table:priorjoint}, respectively. 
We include the `min' case in purple, `mid' case in blue, `max' case in grey, and fiducial case in green.
The fiducial case keeps the \hmcode feedback amplitude $\log{B}$ fixed at the DM-only value of 0.496 and does not include any systematic uncertainties. 
We further include the Planck CMB contour for comparison in red.
Right: Same as the left panel, except now showing contours in $\Omega_{\mathrm m}$ against $\sigma_8 \Omega_{\mathrm m}^{0.5}$, orthogonal to the $\sigma_8 - \Omega_{\mathrm m}$ degeneracy direction.
}
\label{figjoint}
\end{figure*}

\subsection{Including Photometric Redshift Uncertainties}
\label{photsec}

We now turn to the third key systematic: photometric redshift uncertainties. We consider 7 tomographic bins in our analysis, and therefore introduce 7 new parameters to allow the source distribution of each tomographic bin to shift along the redshift axis, preserving the shape of each distribution.

As shown in Table~\ref{table:prisys}, we consider two distinct cases for our priors, 
one where $-0.1 < {\Delta}z_i < 0.1$ for each tomographic bin (varied uniformly), 
and one where Gaussian priors are obtained from \citealt{choi15} (with minor variations; hereafter also denoted C15) for the first five tomographic bins, while the last two bins are varied uniformly between -0.1 and 0.1. The informative priors are derived by fitting angular cross-correlation function measurements between sources in each tomographic bin and an overlapping spectroscopic sample 
from BOSS, as detailed in \citet{choi15}.
The priors are only available for the first five tomographic bins given the redshift coverage of BOSS
and the numbers differ slightly from those presented in \citet{choi15} because of more conservative error estimation and a different normalization scheme\footnote{See Section 2.5 of \citet{choi15} for details of the redshift probability distribution shifting procedure. Whenever negative shifts were applied, the \textsc{BPZ} probability distributions were renormalized by the integrated probability including negative redshifts. The renormalization performed in the process of obtaining our priors neglect the probability distribution shifted to negative redshifts.}. The two sets of numbers agree well within the 1$\sigma$ error bars, and the choice of normalization has a negligible impact on our analysis.

In Fig.~\ref{figphotoz}, we show the marginalized posterior contours in the $\sigma_8 - \Omega_{\mathrm m}$ plane for the different treatments of the redshift distribution.
For both informative and non-informative prior cases, we find only small changes in the contours along the $\sigma_8 - \Omega_{\mathrm m}$ plane as compared to the fiducial scenario where the redshift distribution is fixed, in agreement with a similar analysis in \citet{dessv}. Here, the informative contour is marginally expanded, but not in the region that would increase the agreement with Planck, while the non-informative contour is marginally expanded and shifted towards the Planck contour. 

While the tomographic shifts are given by C15 for the informative case (aside from the last two bins, which are found to be entirely unconstrained within the prior), at 95\% CL we find 
$-0.06 < {\Delta}z_1 < 0.1$,
$-0.03 < {\Delta}z_2 < 0.1$, 
$-0.1 < {\Delta}z_3 < 0.1$,
$-0.05 < {\Delta}z_4 < 0.1$,
$-0.1 < {\Delta}z_5 < 0.1$,
$-0.1 < {\Delta}z_6 < 0.06$,
$-0.1 < {\Delta}z_7 < 0.1$
for the case with uniform priors ($\left|{\Delta}z_i\right| = 0.1$). These bounds demonstrate that the constraints are weak and the uncertainties would increase with wider priors.
For the C15 case, we find $\Delta\chi^2_{\rm eff} = 1.5$ and $\Delta{\rm DIC} = 11$, such that the more complex model seems to be strongly disfavored by the data as compared to the fiducial model.
This finding is in agreement with the conclusions of \citet{choi15}, who showed that the best-fitting model for the one-parameter shift of the $n(z)$ used in our `min' case (and advocated in \citealt{dessv}) is actually insufficient to encompass the errors in the CFHTLenS redshift distributions. Significantly better fits to the spectroscopic-photometric cross-correlation clustering measurements can be obtained when the width of the redshift distribution is allowed to vary along with the peak. We will explore this further freedom in the photometric redshift distribution uncertainty in future work.

For the case with uniform priors on all tomographic bins, 
$\Delta\chi^2_{\rm eff} = -6.0$ and $\Delta{\rm DIC} = 0.97$. Thus, the goodness of fit improves when allowing for uniform deviations around the fiducial distribution, as opposed to the shifts advocated by C15, but keeping the new degrees of freedom is not preferred by the data. 

\subsection{Joint Account of Systematic Uncertainties}
\label{jointsec}

Within the $\Lambda$CDM cosmology, we now consider the joint analysis of systematic uncertainties coming from intrinsic alignments of galaxies, nonlinear baryonic physics in the matter power spectrum, and photometric redshift errors. To this end, we set up three distinct cases for the priors on the new degrees of freedom, a minimum (`min') case with informative priors, a maximum (`max') case with conservative priors, and a middle (`mid') case with a combination of informative and conservative priors. We list the priors for these three cases in Table~\ref{table:priorjoint}. Meanwhile, 
for the underlying cosmology, we continue to impose the fiducial priors given in Table~\ref{table:prisys}.

For the `min' case, we assume an informative uniform prior on the intrinsic alignment amplitude of $-6 < A < 6$, and exclude a luminosity or redshift dependence of the intrinsic alignment signal. We assume an informative uniform prior on the $\hmcode$ amplitude $0.3 < \log B < 0.6$, while the fiducial tomographic redshift distributions are perturbed by the shifts given in \citealt{choi15} (aside from the last two bins, as discussed in Section~\ref{photsec}). For the `mid' case, we keep the same settings as for the `min' case, except we now allow for a luminosity dependence via $\beta = 1.13 \pm 0.25$, and we increase the error bars on the C15 redshift shifts to $\left|{\Delta}z_i\right| = 0.05$. Lastly, for the `max' case, we impose wide priors on all systematic degrees of freedom, such that $-50 < \{A, \eta, \beta\} < 50$, $0 < \log B < 2$, and $-0.1 < {\Delta}z_i < 0.1$.

As described in Section~\ref{measlab}, our measurements are evaluated at 7 angular bins (for each of the 7 tomographic bins). However, for the `max' case, we consider removing the dependence on nonlinear scales in the matter power spectrum altogether. To this end, we follow \citet{planckmg15} and cut our data vector by removing $\xi_-$ entirely and keeping $\xi_+$ for $\theta > 17~{\rm arcmins}$. In practice, this implies we only keep our measurements of $\xi_+$ at 21.7~arcmins and 43.0~arcmins (since we already remove the $\xi_+$ measurements at 85.2~arcmins in the fiducial data vector), such that the downsized data vector consists of 56 elements (from the fiducial vector of 280 elements, itself originally downsized from 392 elements).

In Fig~\ref{figjoint}, we show the marginalized posterior contours for the three systematic cases along the $\sigma_8 - \Omega_{\mathrm m}$ plane. For the `min' case, which includes informative priors on the systematic uncertainties, the main change to the contour comes from the freedom in the intrinsic alignment amplitude. This is because the impact of the baryonic and photometric redshift uncertainties is marginal when imposing informative priors (as seen in Sections~\ref{barsec}~and~\ref{photsec}). As expected from the left panel of Fig.~\ref{figia}, where only the intrinsic alignment amplitude is varied freely (in addition to the vanilla cosmological parameters), the expanded `min' contour primarily shifts towards smaller values of the matter density, and away from the Planck contour, as compared to the fiducial case with no systematic uncertainties included. 
Thus, for the min case, $\sigma_8 \Omega_{\mathrm m}^{0.5} = 0.372^{+0.023}_{-0.022}$ (68\% CL), as compared to $\sigma_8 \Omega_{\mathrm m}^{0.5} = 0.401^{+0.016}_{-0.017}$ (68\% CL) for the fiducial case. 

Proceeding from the `min' case to the `mid' case, the expanded contour is shifted back to
values of $\{\sigma_8, \Omega_{\mathrm m}\}$ that overlap with those of the fiducial scenario
due to the additional degree of freedom from the luminosity dependence. More specifically, the parameter $\beta$ is sufficiently large to decrease the intrinsic alignment signal, consistent with the behavior seen for the informative case in the right panel of Fig.~\ref{figia}. The contour for the `mid' case is moreover
expanded in the $\sigma_8$ direction due to the photometric redshift uncertainties,
consistent with the behavior seen in Fig.~\ref{figphotoz}. The combination of these two effects brings the
`mid' contour 
in greater agreement with Planck (as compared to the fiducial case), 
as also quantified by $\sigma_8 \Omega_{\mathrm m}^{0.5} = 0.404^{+0.021}_{-0.021}$ (68\% CL). 

In the `max' case, a combination of conservative priors and downsized data vector increases the size of the contour to such an extent that it encloses all of the aforementioned contours, including the Planck contour. It is clear that in this pessimistic scenario, 
the CFHTLenS dataset is only able to place weak constraints in the $\sigma_8 - \Omega_{\mathrm m}$ plane, with $\sigma_8 \Omega_{\mathrm m}^{0.5} = 0.395^{+0.074}_{-0.064}$ (68\% CL),
although it does retain the anti-correlated shape between $\sigma_8$ and the matter density. 
Thus, from the marginalized posterior contours, it seems the introduction of key systematic uncertainties from intrinsic alignments, baryons, and photometric redshifts is able to alleviate the tension with Planck for the more conservative `mid' and `max' cases (as compared to the fiducial case). However, as pointed out in e.g.~\citet{raveri15}, there is a risk of biasing one's conclusions when assessing dataset concordance from marginalized posterior contours. We therefore proceed to evaluate the three cases of the joint analysis more quantitatively.

\begin{table}
\begin{center}
\caption{Exploring changes in $\log \mathcal{C}$, $\log \mathcal{I}$, and $\ln \mathcal{B}_{01}$ for different choices of systematic uncertainties given fiducial cosmological priors. 
For $\log \mathcal{C}$ and $\log \mathcal{I}$, positive values indicate concordance between CFHTLenS and Planck while negative values indicate discordance between the datasets. For $\ln \mathcal{B}_{01}$, which only considers CFHTLenS data, positive values indicate the fiducial model is preferred, while negative values indicate the extended model is preferred.
We note that `$\ln$' refers to the natural logarithm, while `$\log$' refers to the common logarithm (base 10).
The Planck CMB evidence is given by $\ln \mathcal{Z} = -5680.5$, while the vanilla CFHTLenS evidence is given by $\ln \mathcal{Z} = -214.9$ when using \hmcode with feedback amplitude $\log B = 0.496$. The vanilla CFHTLenS evidence for the downsized data vector of the `max' case is $\ln \mathcal{Z} = -47.9$ (considering $\log B = 0.496$).
From these numbers and those in the table, the individual evidences can be reconstructed.
Moreover, Planck's $\rm{DIC} = 11297.1$, which can be used to reconstruct the joint DIC estimates from the $\log {\mathcal{I}}$ estimates in this table and the ${\rm DIC}$ estimates from Table~\ref{table:pridic}.
The errors in our calculations of $\log \mathcal{C}$ and $2 \ln \mathcal{B}_{01}$ are approximately 0.2 and 0.3, respectively. For the joint systematics calculations, we consistently use \hmcode for the nonlinear matter power spectrum.
}
\begin{tabular}{p{3.1cm}>{\raggedleft}p{1.6cm}>{\raggedleft\arraybackslash}p{0.99cm}>{\raggedleft\arraybackslash}p{0.99cm}}
\toprule
Model & $\log {\mathcal{C}}$ & $\log {\mathcal{I}}$ & $2 \ln \mathcal{B}_{01}$ \\
\midrule
vanilla (\hmcode, $B$ fixed) & $-0.47$ & $-0.98$ & $0$ \\
vanilla (\halofit) & $-0.96$ & $-1.6$ & $-1.9$ \\
vanilla + min case & $-1.8$ & $-2.6$ & $-6.6$ \\
vanilla + mid case & $0.51$ & $-0.28$ & $1.3$ \\
vanilla + max case & $0.90$ & $0.62$ & $-0.52$ \\
\bottomrule
\end{tabular}
\label{table:evidence}
\end{center}
\end{table}

In the joint analysis, we find 
$\Delta\chi^2_{\rm eff} = -10$, $\Delta{\rm DIC} = 1.6$, and $2\ln \mathcal{B}_{01} = -6.6$ for the `min' case (with respect to the fiducial case, and where $\mathcal{B}_{01}$ refers to the Bayes factor defined in Section~\ref{modsec}). This illustrates the usefulness of model selection based on multiple statistics, as the DIC and evidence estimates point in somewhat different directions. 
This may be a reflection of the parameter priors, analogous to the `Jeffreys-Lindley paradox' (\citealt{lindley,jeffreys}, also see \citealt{cousins13})
We follow the prescription in \citet{ktp06}, and conclude that despite the increased complexity, the improvement in the evidence is sufficiently large to warrant the `min' case as favored by the data.

For the `mid' case, $\Delta\chi^2_{\rm eff} = -1.1$, $\Delta{\rm DIC} = 12$, and $2\ln \mathcal{B}_{01} = 1.3$, which implies a preference against the more complex model, at a greater significance when employing the DIC as compared to the evidence.
For the `max' case, we find $\Delta\chi^2_{\rm eff} = -25$, $\Delta{\rm DIC} = 19$, and $2\ln \mathcal{B}_{01} = -0.52$, such that this model is roughly equally favored to the `vanilla' model when employing the evidence, but highly disfavored when accounting for its increased complexity. We note that the changes in $\chi^2_{\rm eff}$, evidence, and DIC for the `max' case are when compared to a fiducial case using the downsized data vector keeping `large' angular scales but without systematics. For this fiducial case, $\chi^2_{\rm red} = 1.74$, up from $\chi^2_{\rm red} = 1.51$ when 
including `all' scales, suggesting that the high $\chi^2_{\rm red}$ is exacerbated on large scales (in agreement with the discussion in Section~\ref{chopri}).

Turning to the question of dataset concordance between CFHTLenS and Planck, we computed both the $\log {\mathcal{C}}$ and $\log {\mathcal{I}}$ statistics (defined in equations~\ref{eveqn}~and~\ref{diceqn}, respectively), with results shown in Table~\ref{table:evidence}. These two dataset concordance tests are respectively based on the Bayesian evidence and information theory, detailed in Section~\ref{modsec}. For the scenario without systematic uncertainties included, we find $\log {\mathcal{C}} = -0.47$ and $\log {\mathcal{I}} = -0.98$ when employing \hmcode for the nonlinear matter power spectrum (with DM-only feedback amplitude of $\log B = 0.496$), showing substantial degree of discordance between CFHTLenS and Planck for the two statistics. When instead employing \halofit for the nonlinear matter power spectrum, $\log {\mathcal{C}} = -0.96$ and $\log {\mathcal{I}} = -1.6$, pointing towards strong discordance between the datasets. The increase in the discordance between the datasets when employing \halofit (as compared to \hmcode) is in agreement with the increased separation in the marginalized posterior contours in the $\sigma_8 - \Omega_{\mathrm m}$ plane in Fig.~\ref{figoldnew}. As shown in Table~\ref{table:evidence}, the Bayes factor for the vanilla ${\Lambda}$CDM model with \hmcode relative to \halofit is $2\ln \mathcal{B}_{01} = -1.9$, such that the evidence is marginally improved when employing \halofit as compared to \hmcode (with $\log B = 0.496$).

Including systematic uncertainties, we find $\log {\mathcal{C}} = -1.8$ and $\log {\mathcal{I}} = -2.6$ for the `min' case, which demonstrates strong degree of discordance between the datasets for the $\log {\mathcal{C}}$ statistic and decisive degree of discordance for the $\log {\mathcal{I}}$ statistic. This increase in the discordance between the datasets is consistent with the increased separation between the Planck and CFHTLenS marginalized posterior contours in the $\sigma_8 - \Omega_{\mathrm m}$ plane in Fig.~\ref{figjoint}.
For the `mid' case, we find $\log {\mathcal{C}} = 0.51$ and $\log {\mathcal{I}} = -0.28$, such that $\log {\mathcal{I}}$ is consistent with weak discordance, while $\log {\mathcal{C}}$ lies on the border between weak concordance and substantial concordance. This again seems to be consistent with the partial overlap in the CFHTLenS and Planck marginalized posterior contours in Fig.~\ref{figjoint}.
For the `max' case, we find $\log {\mathcal{C}} = 0.90$ and $\log {\mathcal{I}} = 0.62$, such that $\log {\mathcal{I}}$ is consistent with substantial concordance between the datasets, while $\log {\mathcal{C}}$ lies on the border between substantial concordance and strong concordance. Thus, even though the marginalized posterior contour for this case completely envelopes the Planck contour in Fig.~\ref{figjoint}, the degree of concordance between the datasets is not as impressive as naively expected prior to the execution of the $\log {\mathcal{C}}$ and $\log {\mathcal{I}}$ tests.

We can conclude that the question of dataset concordance between CFHTLenS and Planck is sensitive to the exact details of the systematic uncertainties coming from intrinsic alignments, photometric redshift uncertainties, and baryonic uncertainties in the nonlinear matter power spectrum. For our `min' scenario there seems to strong-to-decisive discordance between the datasets, for the `mid' scenario there seems to be weak discordance to substantial concordance, and for the `max' scenario there seems to be substantial concordance. These results are largely in agreement with the {\it a~priori} expectation from the marginalized posterior contours in the $\sigma_8 - \Omega_{\mathrm m}$ plane. 
For the three joint systematics cases, the `min' case is the one most favored by the cosmic shear data, but it is also the one that shows the greatest degree of discordance with Planck. 
The general agreement between the results from the $\log {\mathcal{C}}$ and $\log {\mathcal{I}}$ tests indicates that one may be able to compute the degree of concordance between datasets more easily from existing MCMC chains for parameter estimation, instead of embarking on new evidence calculations.

While it is more than plausible that CFHTLenS contains unaccounted systematics beyond those considered here (see~\citealt{asgari16}), we note that Planck itself may suffer from internal discordance in its measurements, as pointed out in \citet[but disputed in \citealt{planckinterm}]{addison15}.
Moreover, the discordance between CFHTLenS and Planck CMB temperature bears resemblances to that between the Planck CMB temperature and Planck Sunyaev-Zel'dovich cluster counts \citep{plancksz1, plancksz2}, where the latter is also a probe of the low-redshift universe (as compared to the CMB temperature) and exhibits a similar tension with the Planck CMB temperature in the $\sigma_8 -  \Omega_{\mathrm m}$ plane. 
Although the tension between the Planck observables can be reduced by allowing for a larger uncertainty in the mass bias estimates, 
the cluster count systematics seem to mainly cause shifts along the degeneracy direction, 
and the fiducial cluster count constraint on $\sigma_8 \Omega_{\mathrm m}^{0.3}$ is consistent with CFHTLenS.

\section{Conclusions}
\label{conclusions}

We have revisited the analysis of the CFHTLenS dataset with new cosmic shear measurements and covariance from extensive N-body simulations, along with a new CosmoMC fitting pipeline that accounts for key systematic uncertainties from intrinsic galaxy alignments, baryonic uncertainties in the nonlinear matter power spectrum, and photometric redshift errors. Our data vector comprises 7 tomographic bins covering redshifts up to $z=3.5$, and and 7 angular bins extending to 120~arcminutes. The covariance is constructed from a large suite of 497 N-body simulations, which were $2 \times 2$ sub-divided to 
gain a sufficient number of realizations, preventing our inverse covariance to be unduly biased by noise in the sample covariance estimator.

We used the new measurements and covariance to explore the consistency of cosmic microwave background data from Planck with cosmic shear data from CFHTLenS, 
given increasing degrees of freedom from the systematic uncertainties.
To this end, our CosmoMC pipeline calculates the cosmic shear likelihood and allows for three degrees of freedom for the intrinsic alignments, in the form of an amplitude along with a redshift and luminosity dependence. The pipeline further allows for one degree of freedom for the baryonic effects on nonlinear scales in the matter power spectrum in \hmcode, which we have incorporated as a distinct CosmoMC module that internally communicates with our likelihood module. Lastly, the pipeline allows for seven degrees of freedom for the photometric redshift uncertainties, which are manifested by shifts in each of the tomographic source distributions along the redshift axis, with either uniform or Gaussian priors. Thus, the pipeline allows for a total of 11 nuisance parameters, in addition to the cosmological parameters.

We first applied the pipeline to the measurements considering 4 different sets of cosmological priors, finding the data is not sufficiently powerful to constrain the marginalized posterior contours in the $\sigma_8 - \Omega_{\mathrm m}$ plane without prior-dependence of the results. 
However, the four cases show remarkable agreement along the axis perpendicular to the degeneracy direction, such that the $2\sigma$ tension with Planck effectively remains at the same level of significance regardless of the choice of priors.
We proceeded with the lensing analysis by imposing external priors from local Hubble constant and BBN measurements. As these external priors are completely in agreement with Planck, any further discrepancy between Planck and CFHTLenS with external priors must be coming from CFHTLenS. The new marginalized posterior contours continue to show discrepancy with Planck at the 
$2\sigma$ level.

We then examined if the introduction of systematic degrees of freedom could alleviate the tension between the datasets, and whether any of the extensions are statistically preferred. To this end, we employed the Deviance Information Criterion (DIC), which 
accounts for the Bayesian complexity of models.
We find that a negative intrinsic alignment amplitude, such that $A = -3.6 \pm 1.6$, is preferred by the data at the level of ${\Delta}{\rm DIC} \simeq -5$ as compared to the fiducial model with no systematics included. However, this model seems to be at even greater tension with Planck, 
at the 3$\sigma$ level.
We find that an extension of the intrinsic alignment model to allow for redshift and luminosity dependence brings the relative tension between CFHTLenS and Planck back to its fiducial $2\sigma$ level, but this is because the redshift and luminosity dependence terms allow for values that diminish the intrinsic alignment signal. The three-parameter intrinsic alignment model is marginally disfavored by the data at the level of ${\Delta}{\rm DIC} = 2.4$ with informative priors on the IA parameters, and more strongly disfavored at ${\Delta}{\rm DIC} = 12$ with non-informative priors on the parameters. 

Next, we did not find a preference for nonlinear baryonic physics in the CFHTLenS data, as the extension to allow for a varying amplitude in \hmcode is only favored at ${\Delta}{\rm DIC} = -0.64$ when considering non-informative priors, and disfavored at ${\Delta}{\rm DIC} = 1.4$ when considering informative priors. Allowing for the \hmcode feedback amplitude to account for nonlinear baryonic physics has a marginal impact on the tension between CFHTLenS and Planck.
We moreover allowed for photometric redshift uncertainties by imposing uniform priors of $\left|{\Delta}z_i\right| = 0.1$ for each tomographic bin. Allowing for deviations around the fiducial redshift distributions produces an improvement in the goodness of fit, at the level of $\Delta\chi^2_{\rm eff} = -6.0$, but we find that introducing the new degrees of freedom is not preferred by the data at ${\Delta}{\rm DIC} = 0.97$. We considered a case where the redshift perturbations are obtained from the cross-correlation analysis of \citet{choi15}, but found that these redshifts are even more disfavored, at the level of ${\Delta}{\rm DIC} = 11$. As for the tension between CFHTLenS and Planck, the photometric redshift uncertainties only have a marginal impact.

Thus, when introducing the systematic uncertainties independently, only the intrinsic alignment amplitude is substantially preferred by the CFHTLenS data. However, the negative amplitude may be considered unphysical, likely caused by overly simplistic IA modeling and/or unaccounted systematics, 
and it only increases the tension between the Planck and CFHTLenS datasets. Aside from the question of whether the systematic degrees of freedom are preferred by the data, we also find no strong relief in the tension between the two datasets when independently allowing for baryonic physics on nonlinear scales, photometric redshift uncertainties, and non-minimal extensions to the intrinsic alignment model. 

We moreover considered three distinct cases for the joint account of the systematic uncertainties (detailed in Table~\ref{table:priorjoint}). The first of these cases is the `min' case where we impose an informative prior on the intrinsic alignment amplitude (excluding a possible luminosity or redshift dependence), an informative prior on the \hmcode feedback amplitude, and informative priors on the source redshift distributions given by \citet{choi15}. The second case is the `mid' case, where we impose informative priors on the intrinsic alignment amplitude, redshift, and luminosity dependence. For this case we continue to impose an informative prior on the \hmcode feedback amplitude. We also continue to use the \citet{choi15} shifts in the fiducial redshift distributions, but with error bars given by $\left|{\Delta}z_i\right| = 0.05$. Lastly, the third case is the `max' case, where we impose non-informative priors on all three intrinsic alignment parameters, a non-informative prior on the \hmcode feedback amplitude, and uniform priors on the redshift distributions given by $\left|{\Delta}z_i\right| = 0.1$. For the `max' case, we also only keep `large' scales, by which we remove all $\xi_-$ measurements and only keep $\xi_+$ measurements for $\theta > 17$ arcmins.

For the `min' case, the main imprint comes from varying the intrinsic alignment amplitude, which increases the tension with Planck, analogously to when the parameter is considered independently. 
The goodness of fit significantly improves with $\Delta\chi^2_{\rm eff} = -10$ and the Bayes factor favors the `min' case by $2 \ln \mathcal{B}_{01} = -6.6$ (as compared to the fiducial cosmological model with no systematic uncertainties). Despite the increased complexity, manifested in ${\Delta}{\rm DIC} = 1.6$, the evidence is sufficiently improved to warrant the `min' case as more favored as compared to the vanilla $\Lambda$CDM model.
For the `mid' case, the marginalized posterior contour shows strong overlap with that of the fiducial case, as the luminosity dependence diminishes the intrinsic alignment signal, while the baryonic and photometric redshift errors each contribute to a marginal increase in the area of the posterior contour, bringing CFHTLenS into greater concordance with Planck. 
We find a marginal improvement in the goodness of fit, given by $\Delta\chi^2_{\rm eff} = -1.1$. 
However, the Bayes factor disfavors the `mid' case by $2 \ln \mathcal{B}_{01} = 1.3$ 
and the information criterion disfavors it more strongly by ${\Delta}{\rm DIC} = 12$.

The largest impact on the cosmological constraints comes from the `max' case, where the cutting of angular scales and non-informative priors on the 11 systematic parameters result in cosmological constraints so weak that the marginalized posterior contour for CFHTLenS completely envelopes the Planck contour. For this case, we find a significant improvement in the best fit, at $\Delta\chi^2_{\rm eff} = -25$, while the Bayes factor shows a marginal improvement of $2 \ln \mathcal{B}_{01} = -0.52$ (as compared to a fiducial cosmological model using the downsized data vector). 
However, the increased complexity of the model renders it highly disfavored at ${\Delta}{\rm DIC} = 19$, which highlights the usefulness of considering multiple statistical tools for purposes of model selection.

In more carefully assessing the degree of concordance or discordance between CFHTLenS and Planck, we further employed `data concordance tests' as quantified by $\log {\mathcal{C}}$ and $\log {\mathcal{I}}$ (defined in Section~\ref{modsec}), grounded in the Bayesian evidence and deviance information criterion, respectively.
With these statistical tools, we find strong-to-decisive discordance between the two datasets for the `min' case, as evidenced by $\log {\mathcal{C}} = -1.8$ and $\log {\mathcal{I}} = -2.6$, respectively. We further find weak discordance to substantial concordance for the `mid' case, as evidenced by $\log {\mathcal{I}} = -0.28$ and $\log {\mathcal{C}} = 0.51$, while there is substantial concordance for the `max' case, as evidenced by $\log {\mathcal{C}} = 0.90$ and $\log {\mathcal{I}} = 0.62$. The outcome of these concordance tests generally agree with the {\it a~priori} expectation from the marginalized posterior contours, although the degree of concordance is weaker for the `max' case than naively expected. For the three joint systematics cases, it is interesting to note that the case most discordant with Planck is also the one most favored by the data. We can also conclude that the results from the $\log {\mathcal{C}}$ and $\log {\mathcal{I}}$ data concordance tests generally agree with one another, indicating that either one may be used to assess the degree of concordance between datasets in future analyses.

Our new measurements and fitting pipeline are publicly available at the address \sjaddress.
We have extended the pipeline to account for joint analyses of cosmic shear tomography, galaxy-galaxy lensing, and redshift space distortion measurements, including the full covariance for overlapping surveys. We plan to release this extended pipeline as part of an upcoming paper to constrain modified gravity and neutrino physics (Joudaki et al. in prep).

\section*{Acknowledgements}
We much appreciate useful discussions with Jason Dossett, Marian Douspis, Farhan Feroz, Will Handley, Manoj Kaplinghat, Antony Lewis, Niall MacCrann, Gregory Martinez, and Marco Raveri. We greatly thank Luke Hodkinson for help with parallelizing our code. We also greatly thank Robin Humble and Jarrod Hurley for HPC support. We acknowledge the use of \astac time on Swinburne's swinSTAR and NCI's Raijin machines, with special thanks to Amr Hassan and Jarrod Hurley for these resources. We acknowledge the use of \camb and \cosmomc packages (\citealt{Lewis:2002ah}; \citealt{LCL}). 
CB acknowledges the support of the Australian Research Council through the award of a Future Fellowship.
CH and AC acknowledge support from the European Research Council under grant number 240185 and 647112.
BJ acknowledges support by an STFC Ernest Rutherford Fellowship, grant reference ST/J004421/1.
HH is supported by an Emmy Noether grant (No. Hi 1495/2-1) of the Deutsche Forschungsgemeinschaft.
MV acknowledges support from the European Research Council under FP7 grant number 279396 and the Netherlands Organisation for Scientific Research (NWO) through grants 614.001.103.
Computations for the $N$-body simulations were performed on the GPC supercomputer at the SciNet HPC Consortium. SciNet is funded by: the Canada Foundation for Innovation under the auspices of Compute Canada; the Government of Ontario; Ontario Research Fund - Research Excellence; and the University of Toronto.

\bibliographystyle{mn2e}
\bibliography{cosmicshear}

\end{document}